\documentclass[draft, onecolumn]{IEEEtran}
\IEEEoverridecommandlockouts
\usepackage{ifpdf}
\usepackage{cite}
\usepackage{graphicx}
\usepackage{amsmath}
\usepackage{algorithm}
\usepackage{algpseudocode}
\usepackage{array}
\usepackage{blindtext}
\usepackage{epstopdf}
\usepackage{bm}
\usepackage{enumerate}
\usepackage{amsbsy}
\usepackage{amssymb}
\usepackage{amsthm}
\usepackage{amscd}
\usepackage{subfigure}
\usepackage{color}
\usepackage{booktabs}
\usepackage{multirow}
\usepackage{hhline}
\usepackage{makecell}
\usepackage{url}
\usepackage{tabularx}
\usepackage{cellspace}
\usepackage{boldline}
\usepackage[inline]{enumitem}
\usepackage{subfigure}

\usepackage{amsmath}
\usepackage{algpseudocode}
\usepackage{amsthm}
\usepackage{dsfont}
\usepackage{algorithm}
\usepackage{thmtools}
\usepackage{amssymb}
\usepackage[dvipsnames]{xcolor}
\usepackage{todonotes}

\def\ba{\boldsymbol{a}}

\def\bp{\boldsymbol{p}}

\def\bs{\boldsymbol{s}}

\def\bx{\boldsymbol{x}}
\def\by{\boldsymbol{y}}

\def\bR{\boldsymbol{R}}

\def\calC{\mathcal{C}}

\def\calS{\mathcal{S}}

\def\bzero{\boldsymbol{0}}
\def\bone{\boldsymbol{1}}

\newcommand{\binaryB}[1][]{\mathds{B}^{#1}}
\newcommand{\complexC}[1][]{\mathds{C}^{#1}}
\newcommand{\expecE}[1]{\mathds{E}\left\{{#1}\right\}}

\newcommand{\realR}[1][]{\mathds{R}^{#1}}
\newcommand{\realRp}[1][]{\mathds{R}_{+}^{#1}}

\newcommand{\zinv}[1][]{z^{\ifx&#1&-1\else-#1\fi}}

\def\deg{^\circ}
\def\Oh#1{\mathcal{O}\left(#1\right)}

\def\st{~\mathrm{s.t.}~}
\def\ow{~\mathrm{otherwise}}

\def\eg{\textit{e.g.}}
\def\ie{\textit{i.e.}}

\theoremstyle{definition}

\algnewcommand\algorithmicinput{\textbf{Input:}}
\algnewcommand\Input{\item[\algorithmicinput]}
\algnewcommand\algorithmicoutput{\textbf{Output:}}
\algnewcommand\Output{\item[\algorithmicoutput]}
\algnewcommand\algorithmicinit{\textbf{Initialize:}}
\algnewcommand\Init{\item[\algorithmicinit]}

\title{Efficient Waveform Covariance Matrix Design and Antenna Selection for MIMO Radar}

\author{Arindam~Bose$^{*\dagger}$,
	Shahin~Khobahi$^{\dagger}$,
	and~Mojtaba~Soltanalian
	\thanks{%
		$^*$Corresponding author (e-mail: \textit{abose4@uic.edu}).
	} 
	\thanks{%
		$^\dagger$The first two authors contributed equally to this work.
	}
	\thanks{%
		This work was supported in part by U.S. National Science Foundation Grants CCF-1704401 and ECCS-1809225. Parts of this work have been presented at the 53rd Asilomar Conference on Signals, Systems, and Computers, Pacific Grove, CA, USA, November 2019 \cite{asilomar2019bose}.
	}
	\thanks{%
		The authors are with the Department of Electrical and Computer Engineering at University of Illinois at Chicago, Chicago, IL 60607, USA (e-mail: abose4@uic.edu, skhoba2@uic.edu, msol@uic.edu).
	}
}

\begin{document}
\maketitle

\begin{abstract}
	Controlling the radar beam-pattern by optimizing the transmit covariance matrix is a well-established approach for performance enhancement in multiple-input-multiple-output (MIMO) radars.
	In this paper, we investigate the joint optimization of the waveform covariance matrix and the antenna position vector for a MIMO radar system to approximate a given transmit beam-pattern, as well as to minimize the  cross-correlation between the probing signals at a number of given target locations.
	We formulate this design task as a non-convex optimization problem and then propose a cyclic optimization approach to efficiently approximate its solution.
	We further propose a local binary search algorithm in order to efficiently design the corresponding antenna positions.
	We show that the proposed method can be extended to the more general case of approximating the given beam-pattern using a minimal number of antennas as well as optimizing their positions.
	Our numerical investigations demonstrate a great performance both in terms of accuracy and computational complexity, making the proposed framework a good candidate for usage in real-time radar waveform processing applications such as MIMO radar transmit beamforming for aerial drones that are in motion. 
\end{abstract}



\section{Introduction and Prior Works}
	Multiple-input-multiple-output (MIMO) radar refers to a unique radar architecture that employs multiple spatially distributed transmitters and receivers--- an emerging technology in the last two decades, attracting a great deal of interest from researchers in radar signal processing community as well as industry \cite{1316398, 1399141, 4516997, 4350230, 4358016, li2008mimo, 4408448, khobahi2019deepdararwave}.
	Unlike a conventional phased array radar, a MIMO transmitter can transmit a set of arbitrary waveforms orthogonal to each other in order to increase the spatial diversity \cite{4350230, 1599974}. 
	One way to exploit such diversity in MIMO systems is by transmitting orthogonal waveforms, and the echo signals can then be re-assigned to the single transmitter. 
	Thus, from an antenna array of $ M_{T} $ transmitters and $M_{R}$ receivers, a MIMO architecture results in a virtual array of $ M_{T}M_{R} $ elements with enlarged size of virtual aperture which provides additional degrees of freedom to improve the spatial resolution \cite{1291865, 4176505}, immunity to interference \cite{1597550}, and an improved target localization capability \cite{5466526, 5393291, 1703855}.\nocite{khobahi2018optimized}
	The advantages of MIMO radar over traditional phased array radar has inspired researchers to address various associated waveform design problems.
	Among them, is the problem of maximizing the output signal-to-interference-plus-noise ratio (SINR) by jointly optimizing the probing signals and the receive filter coefficients \cite{6649991, 8141978, 6472022}.
	Moreover, the probing waveforms transmitted by a MIMO radar can be designed to approximate a desired beam-pattern, and to further minimize the cross-correlation between the transmitted waveforms at a number of given target locations \cite{4524058, 4567663, 7811203, soltanalian2014single}.
	Here, not only the main focus of this design problem is to control the spatial distribution of the transmit power, but also to improve statistical performance of radar system.
	It is known that the said performance of MIMO radar depends heavily on the cross-correlation beampattern which is completely missing in the phased-array case \cite{4276989}.
	
	\nocite{aa1}
	
	An extensive body of work already exists on designing the covariance matrix of radar transmit waveforms in lieu of designing the waveforms directly; which leads to extra degrees of freedom in the design stage.
	For example, in \cite{4276989}, the authors describe a method to optimize the waveform covariance matrix to approximate the desired beam-pattern and minimize the correlation sidelobes using semidefinite quadratic programming (SQP), while in \cite{4524058} a cyclic algorithm (CA) is proposed to synthesize the constant-modulus waveform matrix to approximate a desired covariance matrix.
	A closed-form covariance matrix design method is described in \cite{6747391} to achieve the desired beam-pattern based on discrete Fourier transform (DFT) coefficients and Toeplitz matrices.
	An extension of the DFT-based methods to a planar-antenna-array for constant-modulus waveforms design can be found in \cite{7178393} and \cite{7829401}. \nocite{khobahi2018signal}
	Although the DFT-based techniques for matching the transmit beam-pattern benefit from a lower computational complexity, the performance is not satisfactory for small number of antennas.
	Two algorithms are described in \cite{5765721} to synthesize the waveform covariance matrix for a given desired beam-pattern.
	In the first algorithm, the elements of a square-root matrix of the covariance matrix are parameterized using the coordinates of a hypersphere in order to implicitly optimize the designed square matrix as a positive semidefinite matrix in an iterative manner.
	In the second algorithm, the constraints and redundant information in the covariance matrix are exploited to find a closed-form solution, which although may yield a `pseudo'-covariance matrix, the outcome is not guaranteed to be positive semidefinite.
	For a further study on transmit beam-pattern synthesis approaches, we refer the interested readers to consult \cite{5962371, 7915123, 6698378, 7955071, 7027831, 8378710}, and the references therein.
	
	Note that all the aforementioned algorithms consider only a uniform linear array (ULA) with half-wavelength inter-element spacing, while designing the covariance matrix of the probing signals to match the given transmit beam-pattern. 
	However, it was shown in \cite{8378710} that the selection of the array position can introduce additional degrees of freedom for designing transmit beam-pattern. 
	Namely, by carefully choosing the position of antennas, one can design the desired beam-pattern using much less number of antennas.
	In other words, one can achieve a similar beam-pattern by carefully redistributing the available antennas in a wider transmit field which amounts to increased virtual aperture.
	As a result, a joint optimization of the covariance matrix and the antenna selection vector can achieve superior results compared with existing methods using ULA with the same number of antennas.
	In \cite{8378710}, authors describe a method based on the Alternating Direction Method of Multipliers (ADMM) \cite{boyd2011distributed} to design the antenna selection vector.
	However, a convex relaxation is used to approximate the solution which is not guaranteed to produce an optimal outcome to the non-convex problem (that is NP-hard in general).
	
	\textit{Contributions:} In this paper, we tackle the aforementioned problems using an iterative greedy local search approach inspired by dynamic programming and evolutionary algorithms.
	In each iteration, a set of optimization parameter vectors is chosen to be perturbed and the corresponding objective values are calculated.
	The best parameters are then selected to form the population for the next generation, and then the entire procedure is repeated until a stopping criteria based on the original objective function is met.
	%
	The main contributions of this paper can be described as follows:
	\begin{itemize}
		\item A novel cyclic algorithm is proposed in order to jointly design the covariance matrix of the transmit waveforms and antenna selection vector. 
		The proposed method further allows for minimizing the cross-correlation between the probing signals at a number of given target locations.
		
		\item We design the antenna selection vector using a novel greedy search framework for binary variables. 
		We show that by using the all-one vector as initialization, the proposed algorithm can provide a good approximate solution in a specific number of iterations. 
		Our new framework may be of interest on its own as a general non-convex solver for waveform design in MIMO radar systems with practical constraints.
		
		\item We further provide an extension to the general antenna selection scenario where the algorithm selects the minimum number of antennas.
	\end{itemize}
	To promote reproducible research, the codes for generating the results presented are made publicly available along with this paper.

	\textit{Organization of the paper:} The remainder of the paper is organized as follows. 
	Sections \ref{sec:model} and \ref{sec:prob} describe the general signal model and problem formulation for jointly designing the covariance matrix of the probing signals and the antenna position vector.
	In Section \ref{sec:opt}, we propose a novel cyclic optimization approach to tackle the aforementioned problem, while in Section \ref{sec:evol}, we discuss the antenna selection strategy using an iterative greedy search algorithm in detail.
	We extend the antenna selection problem to a more general case in Section \ref{sec:exten} using a minimal number of antennas.
	Section \ref{sec:num} lays out several numerical examples for the proposed framework.
	Finally, Section \ref{sec:con} concludes the paper.
	
	\textit{Notation:} We use bold-lowercase and bold-uppercase letters to represent vectors and matrices, respectively. $x_i$ denotes the $i$-th element of the vector $\bx$. The superscripts $(\cdot)^*$, $(\cdot)^T$, and $(\cdot)^H$ represent the conjugate, the transpose, and the Hermitian operators, respectively. $\bone_M$ and $\bzero_M$ are the all-one and all-zero vectors of length $M$, respectively. $\binaryB[M]_N = \{\bx~|~\|\bx\|_1 = N, \bx \in \{0,1\}^M, N \leq M\}$ is the set of all binary vectors with size $M$ and $N$ non-zero elements and $\calS^{M}$ is the set of all real symmetric matrices of size $M\times M$. The sets of $ N\times N $ real, real non-negative and complex matrices are denoted by $\realR{}$, $\realRp{}$, and $\complexC{}$, respectively. The $\ell_1$-norm is represented by $\|\cdot\|_1$. $ \Re(\cdot)$ is the (element-wise) real-part of the complex argument. Finally, $\odot$ denotes the Hadamard product of matrices. \nocite{8683876}
	
\section{Signal Model} \label{sec:model}
\begin{figure}
		\centering
		\includegraphics[draft=false, width=0.4\textheight]{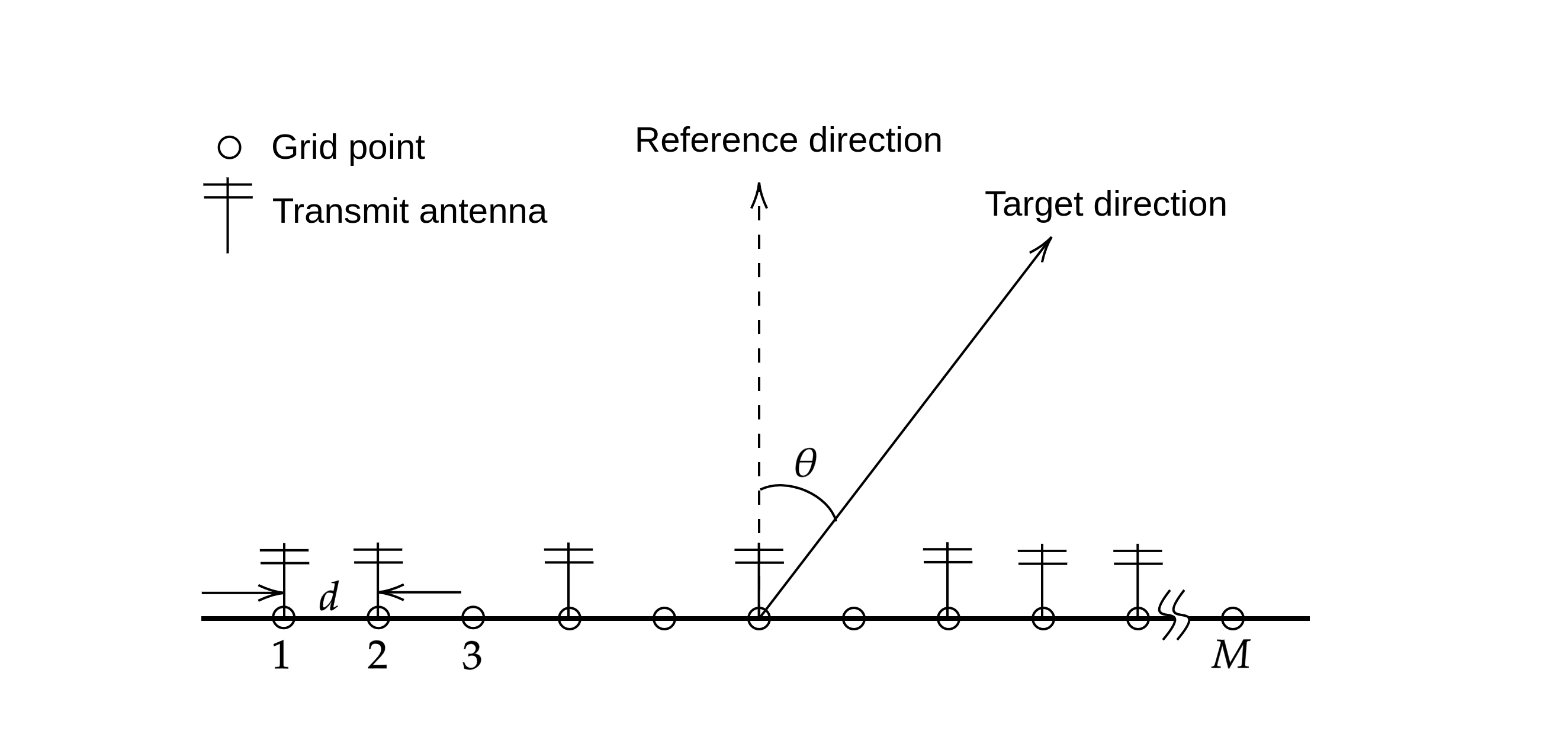}
		\caption{Geometry of a colocated MIMO radar with $M$ grid points with inter-spacing $d$. Only $N$ grid points can be used for antenna placement.}
		\label{fig:radar_scema}
	\end{figure}

	We consider the problem of selecting $N$ transmit antennas placed on a linear array positions with $M (\geq N)$ grid points with equal grid spacing $d$, in order to achieve a desired beam-pattern as depicted in Fig. \ref{fig:radar_scema}.
	A  generalized version of the problem requires choosing the minimum number of antenna positions out of $M$ grid points for the similar purpose.
	In the subsequent sections we consider both scenarios in a detailed manner.
	Let us consider a binary antenna position vector to represent the antenna configuration, viz.
	\begin{align}
		\bp = & [p_1, p_2, \cdots, p_M]^T, \nonumber\\
		&\qquad p_m \in \{0,1\},~m \in \{1, \cdots, M\},
	\end{align}
	where $p_m=1$ indicates that the $m$-th grid point is chosen for antenna placement; otherwise, we have $p_m=0$.
	
	We consider a MIMO radar system transmitting distinct waveforms from each transmission antenna to achieve a desirable beam-pattern.
	Let $s_m(l),$ with $m \in \{1, \cdots, M\}$ and $l \in \{1, \cdots, L\}$, denote the transmit signal from $m$-th antenna, where $L$ is the signal length in discrete-time.
	Assuming that the transmit waveforms are narrow-band and that the propagation is non-dispersive, the baseband waveform at the desired target location $\theta$ can be expressed as \cite{4350230}
	\begin{align}
	\sum\limits_{m=1}^{M} {e^{-j\frac{2\pi}{\lambda} md\sin(\theta)}} s_m(l) \triangleq \ba^H(\theta)\bs(l),\quad l \in \{1,\dots,L\},
	\end{align}
	where $\lambda$ is the wavelength of the transmitted signal, and $\bs(l) = [s_1(l), s_2(l), \cdots, s_M(l)]^T$ is the space-time transmit waveform with length $ M $, and $\ba(\theta)$ is the \textit{steering vector} of the ULA at the direction $\theta$, defined as 
	\begin{align}
		\ba(\theta) = [1, e^{j\frac{2\pi}{\lambda} d\sin(\theta)}, \cdots, e^{j\frac{2\pi}{\lambda} (M-1)d\sin(\theta)}]^T.
	\end{align}
	
	We seek to select $N$ antennas out of $M$ grid point to design the desired beam-pattern. Let $\bp\in\binaryB[M]_N$ denote the antenna selection vector. The corresponding waveform at the target location at the direction $\theta$ with respect to (w.r.t.) the ULA is then given by,
	\begin{align}
		x(l) = (\bp \odot \ba(\theta))^H\bs(l), \qquad l \in \{1,\cdots,L\}.
	\end{align}
	Consequently, the power produced by the waveforms at a generic direction $\theta$ can be written as
	\begin{align} \label{eq:power}
		P(\theta) &= \expecE{|x(l)|^2} \\
		&= (\bp \odot \ba(\theta))^H \expecE{\bs(l)\bs^H(l)} (\bp \odot \ba(\theta)) \nonumber \\
		&= \bp^T \bR \odot \left(\ba(\theta) \ba^H(\theta)\right)^*\bp \nonumber,
	\end{align}
	where
	\begin{align}  
		\bR = \expecE{\bs(l)\bs^H(l)}
	\end{align}
	is the time-averaged covariance matrix of the transmit waveforms $\{\bs(l)\}$.
	As usual in the literature, we refer to the spatial power spectrum defined in \eqref{eq:power}, as the \textit{transmit beam-pattern}.
	Note that, in a similar manner, one can define the cross-correlation terms between the probing signals at locations $\theta$ and $\bar{\theta}$ as
	\begin{align}
		\bar{P}(\theta, \bar{\theta}) \triangleq \bp^T\Re\left\lbrace \bR \odot \left(\ba(\theta) \ba^H(\bar{\theta})\right)^*\right\rbrace\bp.
	\end{align}
	Our goal is to jointly design the antenna selection vector $\bp$ and the covariance matrix $\bR$ of the transmitted waveforms in order to generate the desired beam-pattern while reducing the cross-correlation terms.
	Once $\bR$ has been determined, a signal sequence $\bs(l)$ can be designed that has $\bR$ as its covariance matrix \cite{4276989, 4516997}. 
	
	
\section{Problem Formulation} \label{sec:prob}
	The probing signals transmitted by the MIMO radar system can be designed to enable the system to approximate a desired transmit beam-pattern as well as to minimize the cross-correlation of the signals backscattered from various targets.
	Let $\phi(\theta)$ denote the desired transmit beam-pattern, and $\{\theta_k\}_{k=1}^{K}$ be a grid of points that covers the radial sectors of interest.
	We assume that the said grid comprises of points which are good approximations of the locations of $\tilde{K}$ targets of interest that we wish to probe at locations $\{\theta_k\}_{k=1}^{\tilde{K}}$.
	In addition, we assume that some partial information regarding the target positions are available at hand, \ie, we possess some initial estimates $\{\tilde{\theta}_k\}_{k=1}^{\tilde{K}}$ of $\{\theta_k\}_{k=1}^{\tilde{K}}$.
	In practice, one can obtain $\{\tilde{\theta}_k\}_{k=1}^{\tilde{K}}$ using the Capon spatial spectrum and the generalized likelihood ratio test (GLRT) function  for target localization.
	For example, we form the desired beam-pattern by using the dominant peak locations of the GLRT pseudo-spectrum, denoted by $\{\tilde{\theta}_k\}_{k=1}^{\hat{K}}$ with $\hat{K}$ being the resulting estimate of $\tilde{K}$, as follows:
	\begin{align}
		\phi(\theta) = \left\lbrace
			\begin{array}{ll}
				1, & \theta \in [\tilde{\theta}_k-\frac{\triangle}{2},                \tilde{\theta}_k+\frac{\triangle}{2}],~~ k \in \{1,\cdots,\hat{K}\}, \\
				0, & \ow,
			\end{array}
			\right.
	\end{align}
	where $\triangle$ is the chosen beam-width for each target ($\triangle$ should	be greater than the expected error in $\{\tilde{\theta}_k\}$); see \cite{4350230}.
	
	Our goal is to design $\bR$ such that the transmit beam-pattern $P(\theta)$, approximates the desired beam-pattern $\phi(\theta)$ over the radial sectors of interest in a least squares (LS) sense, and moreover, such that the contribution from all cross-correlation terms $\bar{P}(\theta, \bar{\theta})~(\text{for}~\theta \neq \bar{\theta})$, are minimized (again, in an LS sense) over the set of possible target locations $\{\tilde{\theta}_k\}_{k=1}^{\tilde{K}}$.
	Formally, we make use of the following cost function that incorporates the aforementioned criteria as follows \cite{4350230}:
	\begin{align} \label{eq:J}
		&J(\bp, \bR, \alpha) \\
		&~~ = \frac{1}{K}\sum_{k=1}^{K}{w_k\left|\bp^T \bR \odot \left(\ba(\theta_k) \ba^H(\theta_k)\right)^*\bp - \alpha \phi(\theta_k)\right|^2} \nonumber\\
		&~~ + \frac{2\omega_c}{\tilde{K}(\tilde{K} - 1)} \sum_{p=1}^{\tilde{K}-1}{\sum_{q=p+1}^{\tilde{K}} {\left| \bp^T\Re\left\lbrace \bR \odot \left(\ba(\tilde{\theta}_p) \ba^H(\tilde{\theta}_q)\right)^*\right\rbrace\bp \right|^2}} \nonumber
	\end{align}
	where $\alpha > 0$ is a scaling factor to be optimized, $ \omega_k \geq 0$ is the weight factor for the $k$-th grid point (for $k = 1,\cdots,K$), and $ \omega_c \geq 0 $ is the weight factor for the cross-correlation terms.
	Note that we introduce $\alpha$ as a design parameter in order to achieve the desired transmit beam-pattern that approximates an appropriately scaled version of $ \phi(\theta) $ to take into account different transmit energy  allocations.
	
	In the sequel, we formulate the problem of designing beam-pattern with low cross-correlation for a MIMO radar system as a constrained optimization problem and further impose proper constraints for designing $\bR$ and $\bp$. 
	First, one should impose the constraint that the designed matrix $\bR$ must be positive semi-definite since it is a covariance matrix. Next, under a uniform elemental power constraint, all the diagonal elements of $\bR$ must attain the same value as all antennas are required to transmit uniform power. 
	Hence, the feasible region for the desired transmit covariance matrix can be compactly expressed as,
	\begin{subequations}
		\begin{align}
			&\bR \succeq \bzero, \\
			&R_{mm} = \frac{c}{M},~~\text{for}~m = 1, \cdots, M,
		\end{align}
	\end{subequations}
	with given $c>0$, and $R_{mm}$ denoting the $m$-th diagonal element of $\bR$.
	In the case of designing unimodular sequences, one can simply set $c=1$.
	
	Secondly, due to the fact that we are placing only $N$ antennas in $M (\geq N)$ grid points to achieve the desired beam-pattern, we further impose the constraint that the binary antenna selecting vector $\bp$ should contain $N$ non-zero elements. 
	More precisely, we aim to design $\bp$ according to the following constraints,
	\begin{subequations}
		\begin{align}
			&\|\bp\|_1 = N, \\
			&p_m = \{0,1\}, ~\text{for}~m = 1, \cdots, M,
	\end{align}
	\end{subequations}
	or, equivalently $\bp \in \binaryB[M]_N$. 
	Therefore, the overall transmit covariance optimization problem can be formulated as
	\begin{subequations} \label{eq:prob}
		\begin{align}
			\min_{\bp,\bR, \alpha}\qquad &J(\bp,\bR,\alpha)\\
			\st\qquad & \bR \succeq \bzero, \\
			& R_{mm} = \frac{c}{M},~~\text{for}~m = 1, \cdots, M,\\
			& \|\bp\|_1 = N, \label{eq:probd}\\
			& p_m = \{0,1\}, ~~~\text{for}~m = 1, \cdots, M,\\
			& \alpha > 0.
	\end{align}
	\end{subequations}
	It is easy to verify that the optimization problem in (\ref{eq:prob}) can be categorized as a mixed Boolean-nonconvex problem, especially due to the constraints imposed on $\bp$, the likes of which is very difficult and computationally expensive to solve. In the next subsection, we propose an efficient and novel \emph{cyclic} optimization approach based on semi-definite programming and a greedy search algorithm to tackle the non-convexity of the said problem in \eqref{eq:prob}.
	
\subsection{Cyclic Optimization Algorithm} \label{sec:opt}
	Hereafter, we address the problem of designing the desired covariance matrix $\bR$, the scaling factor $\alpha$, and the corresponding antenna selection vector $\bp$ according to the objective function $J$ and by proposing an alternating optimization approach to tackle the problem of \eqref{eq:prob}. 
	Specifically, the minimization of $J(\bp,\bR,\alpha)$ in \eqref{eq:prob} can be tackled via employing a \emph{cyclic optimization} approach with respect to the design variables $(\bR,\alpha)$ and $\bp$. 
	
    \vspace{7pt}
    $\bullet$ \textbf{Optimization of $\bR$ and $\alpha$:} \label{subsec:optofR}

		For a fixed $\bp$, the minimization problem in \eqref{eq:prob} with respect to $(\bR,\alpha)$ can be recast as
		\begin{subequations} \label{eq:16}
			\begin{align}
				\min_{\bR,\alpha}\qquad &J(\bp,\bR,\alpha)\\
				\st\qquad & \bR \succeq \bzero,\\
				& R_{mm} = \frac{c}{M},~~\text{for}~m = 1, \cdots, M,\\
				& \alpha > 0.
			\end{align}
		\end{subequations}
		Interestingly, it was shown in \cite{4276989} that the above minimization problem with respect to design variables $(\bR,\alpha)$ is convex and can be reformulated as a semi-definite program (SDP), which can then be efficiently solved using numerical methods (\eg, interior point method \cite{nocedal2006numerical}).
	
	\vspace{7pt}
    $\bullet$ \textbf{Optimization of $\bp$:}
    
		On the other hand, for fixed $(\bR,\alpha)$ the optimization problem of \eqref{eq:prob} with respect to the antenna selection vector $\bp$ can be expressed as
		\begin{subequations} \label{eq:17}
			\begin{align}
				\min_{\bp}\qquad &J(\bp,\bR,\alpha)\\ 
				\st \qquad &\bp\in \binaryB[M]_N,
			\end{align}
		\end{subequations}
		where $M$ and $N$ denote the total number of grid points and the total number of antennas we are restricted to choose, in order to form the desired beam-pattern, respectively.
		Note that the constraint set $\binaryB[M]_N$ is not convex due to the (discrete) Boolean constraint of $\bp\in\{0,1\}$ imposed on the antenna selection vector. 
		Put differently, we are interested in minimizing the objective function $J$ over a subset of vertices of a hypercube of dimension $M$, which is represented by $\binaryB[M]_N$.
		We tackle this problem using a greedy search algorithm which is discussed in Section \ref{sec:evol} in detailed manner. 
		
		
		Finally, as mentioned earlier, the cyclic optimization method alternates between the following optimization problems at each cycle:
		\begin{align}\label{eq:18}
			\left(\bR^{(t)},\alpha^{(t)}\right) = \arg \min _{\bR,\alpha} \qquad &J(\bp^{(t-1)},\bR,\alpha)\\
			\st\qquad & \bR \succeq \bzero, \nonumber\\
			& R_{mm} = \frac{c}{M}, \nonumber \\&\qquad \text{for}~m = 1, \cdots, M, \nonumber \\
			& \alpha > 0, \nonumber
		\end{align}
		and
		\begin{align}\label{eq:19}
			\bp^{(t+1)} = \arg\min_{\bp}\qquad  & J(\bp,\bR^{(t)},\alpha^{(t)}) \\
			 \st \qquad & \bp\in\binaryB[M]_N, \nonumber
		\end{align}
		where $t$ denotes the iteration index of the cyclic optimization method.
		

\section{The Proposed Antenna Position Design Technique} \label{sec:evol}
	In this section, we develop a heuristic optimization approach inspired by the dynamic programming and genetic algorithms (a special case of evolutionary optimization technique \cite{DaRonco2014}) equipped with a simple local search to tackle the non-convexity of \eqref{eq:17}.
	Note that the objective function $J(\bp,\bR,\alpha)$ is \emph{quartic} with respect to the vector $\bp$, and thus, it is deemed extremely difficult to solve.
	The ref \cite{8378710} proposes one approach \eqref{eq:17} based on a relaxation of the Boolean constraint (\eg, via the linear relaxation of $\bzero\leq\bp\leq\bone$), which yields a suboptimal solution in expense of heavy computation.
	In this paper, we resort to a greedy search algorithm which can solve the exact problem in \eqref{eq:17} in an efficient manner.
	
	Especially we mimic the process of natural selection for solving an optimization process by iteratively improving the generated set of feasible solutions.
	The fitness of each feasible solution is usually governed by an objective function.
	Then, according to a predefined criteria, the algorithm maintains the best subset of feasible solutions at each iteration to generate better solution \textit{individuals} accordingly.
	Here in each generation, we produce the set of feasible solutions and select the best individual according to a greedy policy, however by design, the particular choice of policy allows for shrinking the cardinality of feasible set in each generation.
	%
	In the following, we go through the main ingredients of the proposed method in order to design the antenna position vector $\bp$.
	
	\subsection{Generation of Feasible Solutions Set}
	As mentioned earlier that our search space for a solution is a subset of vertices of an $M$-dimensional hypercube represented by $\binaryB[M]_N$.
	Hence we undertake a deterministic strategy for the generation of feasible solutions set.
	Note that the binary vector $\bp$ of length $M$ represents a hypercube with $2^M$ vertices. 
	Given the most fitted solution (parent solution) at iteration $k$, \eg, $\bp^{(k)}$, we generate a new set of feasible (candidate) solutions (\ie, offspring of the parent solution) $\bp^{(k+1)}_{\text{CS}}$ as follows:
	\begin{align}
		\bp^{(k+1)}_{\text{CS}} = \left\{\bp \,\,\,\,|\,\,\,H\left(\bp,\bp^{(k)}\right) = 1,\,\, \|\bp\|_1<\|\bp^{(k)}\|_1   \right\},
	\end{align}
	where $H(\bx,\by)$ denotes the Hamming distance between the two vectors, and is defined to be the number of positions $i$ such that $x_i \neq y_i$, where the subscript $i$ denotes the $i$-th element of the corresponding vector.
	In other words, given a parent solution $\bp^{(k)}$, the new set of candidate solutions (CS) is generated as the set of vectors which only differs from $\bp^{(k)}$ in one bit (with one less non-zero element only).
	Then each candidate solution is mutated using a predefined probability (\texttt{prob\_mut}), meaning one randomly selected bit (using uniform sampling) is toggled with the said probability.
	The purpose of mutation is to introduce diversity into the candidate solution set.
	Mutation operators are used in an attempt to avoid local minima by preventing the active bits of candidates (chromosomes) from becoming too similar to each other, thus slowing or even stopping convergence to the global optimum. This reasoning also leads to avoid only taking the fittest of the candidates in generating the next generation, but rather selecting a random (or semi-random) set with a weighting toward those that are fitter \cite{DaRonco2014}.
	Hence, at each iteration the cardinality of the new candidate solution is upper bounded by $\left|\bp_{\text{CS}}^{(k+1)}\right|\leq\|\bp^{(k)}\|_1$.
	This procedure is summarized in Algorithm \ref{alg:generateChildren}.
	
	\begin{algorithm}[t]
		\caption{for generating children set of $\bp$ which are not in \texttt{seen\_children} using mutation}\label{alg:generateChildren}
		\begin{flushleft}
		\begin{algorithmic}[1]
			\Procedure{GenerateChildren}{$\bp$, \texttt{seen\_children}, \texttt{prob\_mut}}
			\State \texttt{children}\textsc{.push}($\bp$)
			\For{$i=1,2,\cdots,$ \textsc{len}($\bp$)}
    			\If{$\bp_i = 1$}
        			\State \texttt{child} $\gets$ Toggle $i$-th bit of $\bp$
					\If{\texttt{rand(1)}$\leq$ \texttt{prob\_mute}}
						\State \texttt{child} $\gets$ Toggle uniformly selected one bit of \texttt{child} with probability \texttt{prob\_mut}
					\EndIf
        			\If{\texttt{child} \textbf{is not in} \texttt{children} and \texttt{seen\_children}}
        			    \State \texttt{children}\textsc{.push}(\texttt{child})
        			\EndIf
    			\EndIf
			\EndFor
			\State \textbf{return} \texttt{children}
			\EndProcedure
			
		\end{algorithmic}
		\end{flushleft}
	\end{algorithm}
	


	\subsection{Selection of the Fittest Solution}
		The goal of selection procedure is to propagate the fittest candidate solution, \ie, the one with the highest fitness value, or in other words lowest objective value, to have a higher probability of generating new offspring or CS for the next iteration (generation) of the algorithm.
		There exist several stochastic and deterministic methods in the literature for the selection procedure, and in this paper, we consider a deterministic approach.
		For fixed $(\bR, \alpha)$, let us denote the objective function \eqref{eq:J} as $J(\bp)$. 
		Having the current CS $\bp_{\text{CS}}^{(k)}$ at hand, we select the fittest solution $\bp^{(k)}$ to be considered for generating new candidate solutions at the next stage as follows:
		\begin{equation}\label{eq:21}
			\bp^{(k)} = \arg\min_{\bp\in\bp_{\text{CS}}^{(k)}}{J(\bp)}.
		\end{equation}
		Next, $\bp^{(k)}$ is used as the seed for generating new CS in the crossover procedure for the next stage of the algorithm.
		This procedure is summarized in Algorithm \ref{alg:bestChildren}.
		\begin{algorithm}[t]
			\caption{for choosing the best child of $\bp$}\label{alg:bestChildren}
			\begin{flushleft}
			\begin{algorithmic}[1]
				\Procedure{BestChild}{\texttt{children}}
				\State Calculate the functional values of $J(\texttt{child}, \bR, \alpha)$ in \eqref{eq:J} for each \texttt{child} in \texttt{children}
				
				\State \textbf{return} the \texttt{child} for which the functional value is minimum
				\EndProcedure
			\end{algorithmic}
			\end{flushleft}
		\end{algorithm}
	
	\subsection{Stopping Criteria}
		Once the selection procedure selects a vector $\bp^{(k)}$ as its output such that $\bp^{(k)}\in\binaryB[M]_N$ or equivalently $\|\bp^{(k)}\|_1=N$, then one can easily argue that a suboptimal solution is obtained. 
		Note that $\bp^{(k)}\in\binaryB[M]_N$ implies $\bp^{(k-1)}\in\binaryB[M]_{N+1}$.
		Hence, one can conclude that if $\bp^{(k)}\in\binaryB[M]_N$, then $\bp^{(k)}$ is a local optimal point in a 1-Hamming distance neighborhood of $\bp^{(k+1)}$ such that $\|\bp^{(k)}\|_1 < \|\bp^{(k-1)}\|_1$, and that $\bp^{(k-1)}\in\binaryB[M]_{N+1}$.
		Moreover, the cardinality of the search space in the 1-Hamming distance local search in \eqref{eq:21} is at most $\|\bp^{(k-1)}\|_1$ 
		and as a result the search space is reduced in each generation.
		The corresponding search process is summarized in Algorithm \ref{alg:choosep}.
		
		\begin{algorithm}[t]
			\caption{for updating \texttt{seen\_children}}\label{alg:updateSeenChildren}
			\begin{flushleft}
			\begin{algorithmic}[1]
				\Procedure{UpdateSeenChildren}{\texttt{seen\_children}, \texttt{children}}
				\For{\textbf{each} \texttt{child} \textbf{in} \texttt{children}}
				\If{\texttt{child} \textbf{is not in} \texttt{seen\_children}}
				\State \texttt{seen\_children}\textsc{.push}(\texttt{child})
				\EndIf
				\EndFor 
				\State \textbf{return} \texttt{seen\_children} 
				\EndProcedure
			\end{algorithmic}
			\end{flushleft}
		\end{algorithm}
		
		\begin{algorithm}[t]
			\caption{for choosing the best $\bp$ using greedy search algorithm (The procedure \textsc{UpdateSeenChildren} is described in Algorithm \ref{alg:updateSeenChildren})}\label{alg:choosep}
			\begin{flushleft}
			\begin{algorithmic}[1]
				\Require $\bR, \alpha$, total number of antennas $N$, total number of grid points $M$, \texttt{prob\_mut} 
				\Ensure $\bp \gets \bone_M$, \texttt{seen\_children} $\gets \emptyset$, \texttt{flag} $\gets 1$
				\While{\texttt{flag}} 
				\State \texttt{children} $\gets$ \textsc{GenerateChildren}($\bp$, \texttt{seen\_children}, \texttt{prob\_mut}) 
				
				\State $\bp \gets$ \textsc{BestChild} (\texttt{children})
				
				\State \texttt{seen\_children} $\gets$ \textsc{UpdateSeenChildren} (\texttt{seen\_children}, \texttt{children})
				
				\If{$\|\bp\|_1 = N$}
				\State \texttt{flag} $\gets 0$
				\EndIf
				
				\EndWhile\label{euclidendwhile}
				
				\State \textbf{return} $\bp$ 
			\end{algorithmic}
			\end{flushleft}
		\end{algorithm}
		\begin{figure}[t]
			\centering
			\includegraphics[draft=false, width=0.37\textheight]{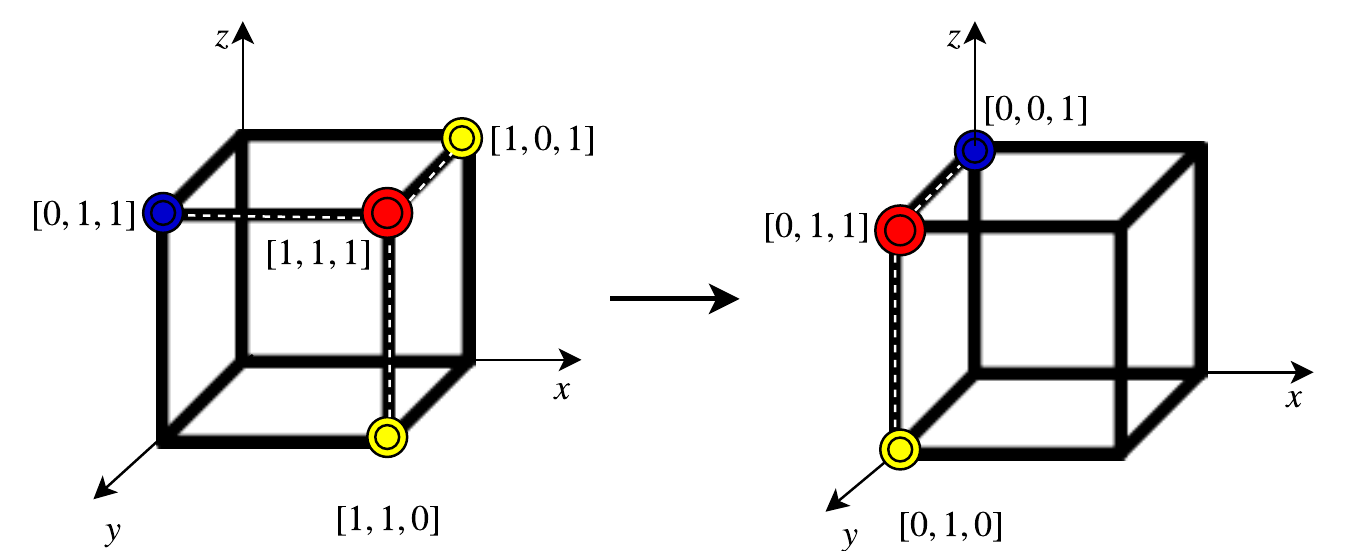}
			\caption{Illustration of the iterations of the proposed greedy search algorithm, where the red vertices denote the parent solution (output of the selection procedure), yellow vertices correspond to the candidate solutions $\bp_{\text{CS}}$, and the blue vertices denote the selected solution for the next iteration.} 
			\label{fig:evolution_final_cube}
			\vspace{-3pt}
		\end{figure}
	
		The above greedy search approach can be best manifested via considering a toy example.
		Assume $M=3$, $N=1$, and the initial antenna position vector $\bp^{(0)}=\bone_M$. 
		Fig. \ref{fig:evolution_final_cube} illustrates the iterations of the proposed greedy search algorithm, where the red vertices denote the parent solution (output of the selection procedure), yellow vertices correspond to the candidate solutions $\bp_{\text{CS}}$, and the blue vertices are the selected solution for the next iteration.
 		At the first iteration, the candidate solutions $\bp_{\text{CS}}^{(1)} = \{[0,1,1],[1,0,1],[1,1,0]\}\subseteq\binaryB[3]_2$, and each member of $\bp_{\text{CS}}^{(1)}$ is also in a 1-Hamming distance of $\bp^{(0)}$.
 		Next, we introduce mutation to each of the candidate solutions according to a predefined mutation probability. Generally, such probability is kept low ( $< 0.5$) to introduce a controlled diversity so that only a small number of candidates are mutated but not all.
 		Let us assume, during the mutation process, only one candidate: $[1,0,1]$ is mutated to $[1,0,0]$, and thus the new candidate set becomes: $\hat{\bp}_{\text{CS}}^{(1)} = \{[0,1,1],[1,0,0],[1,1,0]\}$
 		Next, during the selection procedure, let us assume that the vertex $[0,1,1]$ is chosen as the fittest solution and then used to generate offspring (candidate solutions), \eg $\bp^{(1)}=[0,1,1]$.
 		The new CS generated from $\bp^{(1)}$ is the set $\bp_{\text{CS}}^{(2)} = \{[0,0,1],[0,1,0]\}\subseteq\binaryB[3]_1$.
 		Once again, we apply mutation to all the candidates, however, assume that due to the smallness of the predefined mutation probability, none of the candidates are mutated in this iteration.
 		The fittest solution is then $\bp^{(2)}=[0,0,1]$ and due to the fact that it is a member of the desired set $\binaryB[3]_1$, the algorithm stops. 
 		Next, the antenna position vector $\bp^{(2)}$ is used to design the covariance matrix~$\bR$. 
		
		As it was discussed earlier, we consider the alternating (cyclic) optimization approach to solve the joint optimization of covariance matrix and the antenna position vector.
		Namely, after performing the above greedy search technique for obtaining the solution to \eqref{eq:17} at the $t$-th iteration, \ie obtaining the antenna selection vector $\bp^{(t)}$, we fix $\bp = \bp^{(t)}$ and optimize the objective function with respect to the design variables $(\bR,\alpha)$ according to the method described in Section \ref{subsec:optofR}.
		Finally, the proposed cyclic optimization approach is summarized in Table \ref{tb:fullalgo}.

\section{Extension to General Antenna Selection Scenario} \label{sec:exten}
	So far, we have discussed the joint optimization of transmitted signal covariance matrix and the antenna position while restricted to placing $N$ antennas into $M$ grid points in an optimal manner.
	However, the same proposed greedy search algorithm can be extended to a more general scenario in which we aim to choose the minimum number of antennas $N_{\text{min}}$ for placing in $M$ grid points.
	Namely, in the general antenna selection scenario, we consider the optimization of the signal covariance matrix to form the desired beam-pattern, by letting the algorithm choose the best placement positions while using the \emph{minimum} number of antennas.
	For this general scenario, we consider the following relaxed optimization problem,
	\begin{align}\label{eq:23}
		\min_{\bp,\bR, \alpha}\qquad &J(\bp,\bR,\alpha) + \rho \left(\left|\|\bp\|_1 - N\right|\right) \\
		\st\qquad & \bR \succeq \bzero, \nonumber \\
		& R_{mm} = \frac{c}{M},~~\text{for}~m = 1, \cdots, M, \nonumber\\
		& p_m = \{0,1\}, ~~\text{for}~m = 1, \cdots, M, \nonumber\\
		& \alpha>0, \nonumber
	\end{align}
	where $\rho>0$ denotes the penalty parameter. 
	Note that a lower value of $\rho$ relaxes the solution $\bp$ to have less (than $ N $) number of active antennas by encouraging the total number of non-zero elements $\|\bp\|_1$ of the solution to go far from $N$.
	Conversely, a larger value of $\rho$ keeps the total number of non-zero elements of the solution near $N$.
	Hence, depending on the application, one can choose a lower weight for the total number of active antennas via varying the penalty factor $\rho$. Also, $N$ in \eqref{eq:23} can be interpreted and chosen accordingly as an approximation of the number of antennas one can afford to use. 

	Let $J_2(\bp,\bR,\alpha) \triangleq J(\bp,\bR,\alpha) + \rho \left(\left|\|\bp\|_1 - N\right|\right)$ denote the augmented objective function in \eqref{eq:23}.
	Then, with a slight modification, the same cyclic optimization approach described in Section \ref{sec:opt} can be employed to solve it.
	Note that the extra term in $J_2(\bp,\bR,\alpha)$ only depends on $\bp$, and hence the optimization of $J_2$ with respect to the design variables $(\bR,\alpha)$ remains unchanged and is the same as the procedure described in Section \ref{subsec:optofR}.
	In the previous scenario, we were restricted to a solution $\bp$ such that it satisfies $\bp \in \binaryB[M]_N$.
	However, we have no such restriction in the generalized scenario but only to have $\bp\in\{0,1\}^M$ and instead we are interested in choosing the minimum number of antennas while optimizing their positions. 

	In order to optimize the new augmented objective function $J_2$ with respect to the vector $\bp$, we only need to change the stopping criteria and the fitness function.
	In this general case, the fitness function is considered to be $J_2(\bp,\bR,\alpha)$ and the corresponding stopping criteria for the greedy search approach can be described as follows.
	As it was discussed earlier in Section \ref{sec:evol}, starting with the initialization $\bp^{(0)}=\bone_M$, at each iteration of the proposed search algorithm, the parent node $\bp^{(k)}\in \binaryB[M]_{N-k}$ is a local optimal point in a 1-Hamming distance neighborhood of $\bp^{(k-1)}$. Hence, a heuristic proper stopping criteria can be assumed when the following condition is satisfied at the $k$-th inner iteration of the search process:
	\begin{equation}\label{eq:24}
		H\left(\bp^{(k)},\bp^{(k-1)}\right) = 0.
	\end{equation}
	In other words, the above criteria implies that the solution $\bp^{(k)}$ is a 1-Hamming distance optimal point for its parent $\bp^{(k-1)}$ as well as the newly generated candidate solutions $\bp^{(k+1)}_{\text{CS}}$.

\begin{table}[t]
	\footnotesize
	\caption{The Proposed Joint Optimization Method}
	\label{tb:fullalgo}
	\setlength{\extrarowheight}{5pt} \centering
	\begin{tabular}{p{3.3in}}
		\hline \hline
		\textbf{Step 0}: Initialize the antenna position vector $\bp^{(0)} = \bone_M$, the complex covariance matrix $\bR^{(0)}\in\complexC{N\times N}$, and the scaling factor $\alpha^{(0)}\in \realRp{}$, and the outer loop index $t=1$.\\
		
		\textbf{Step 1}: Solve the convex program of \eqref{eq:18} using the procedure described in Section \ref{subsec:optofR} and obtain $\left(\bR^{(t)},\alpha^{(t)}\right)$.\\
		
		\textbf{Step 2}: Employ the proposed greedy search approach described in Section \ref{sec:evol} and solve the antenna position design program of \eqref{eq:19} to obtain the vector $\bp^{(t+1)}$.\\
		
		\textbf{Step 3}: Repeat steps 1 and 2 until a pre-defined stop criterion is satisfied.\\
		\hline \hline
	\end{tabular}
\end{table} 	

\section{Numerical Examples} \label{sec:num}
	In this section, we provide several examples of numerical simulations in order to assess the performance of our proposed algorithm.
	In the following experiments we assume a colocated narrow-band MIMO radar with a linear array with $ M = 15 $ grid points and half-wavelength inter-grid interval \ie $d = \lambda/2$.
	The range of angle is $ (-90\deg, 90\deg) $ with $ 1\deg $ resolution.
	We set the weights for the $k$-th angular direction as $ w_k = 1$, for $ k=1,\cdots, K $.
	Note that the optimization problem with respect to the variables ($\bR,\alpha$) is carried out using the convex optimization toolbox CVX \cite{grant2014cvx}. 
	Furthermore, we consider the mutation probability as $0.1$.
	\begin{figure}
		\centering
		\includegraphics[draft=false, width=0.40\textheight]{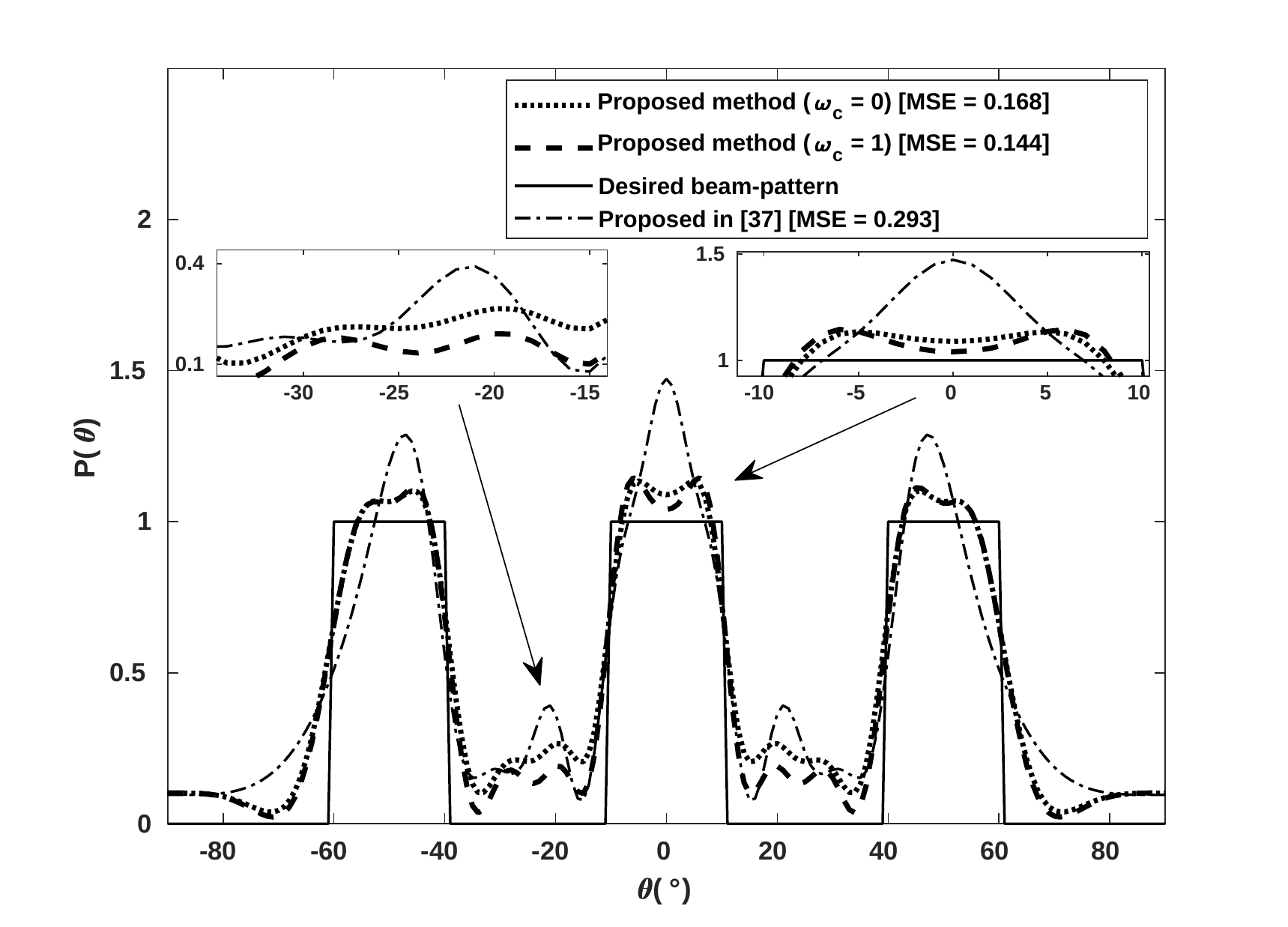}
		\caption{The transmit beam-pattern design for $ M = 15, N = 10 $ with and without the cross-correlation suppression with three mainlobes at $\tilde{\theta} = \{-50\deg, 0\deg, 50\deg\}$ with a beam-width $ \triangle = 20\deg $. It is noticeable that our algorithm outperforms \cite{8378710}.} 
		\label{fig:beampattern_three_lobes}
	\end{figure}

	In Fig. \ref{fig:beampattern_three_lobes}, we consider a design scenario where initial direction of arrival (DoA) information about $\tilde{K} = 3$ targets with unit complex amplitudes, and approximately located at angles $\{-50\deg, 0\deg, 50\deg\}$ is available through the Capon or GLRT method. 
	Hence, we desire to design a symmetric beam-pattern with three directions of interest: $\tilde{\theta}_1 = -50\deg$, $\tilde{\theta}_2 = 0\deg$, and $ \tilde{\theta}_3 = 50\deg $, respectively and the beam-pattern of width $\triangle = 20 \deg$ and thus the given transmit pattern is
	\begin{align*}
		\phi(\theta) = \left\lbrace
		\begin{array}{ll}
			1, & \theta \in [\tilde{\theta}_k-\frac{\triangle}{2}, \tilde{\theta}_k+\frac{\triangle}{2}],~~ k=1,2,3, \\
			0, & \ow.
		\end{array}
		\right.
	\end{align*}
	
	\begin{figure}
		\centering
		\includegraphics[draft=false, width=0.35\textheight]{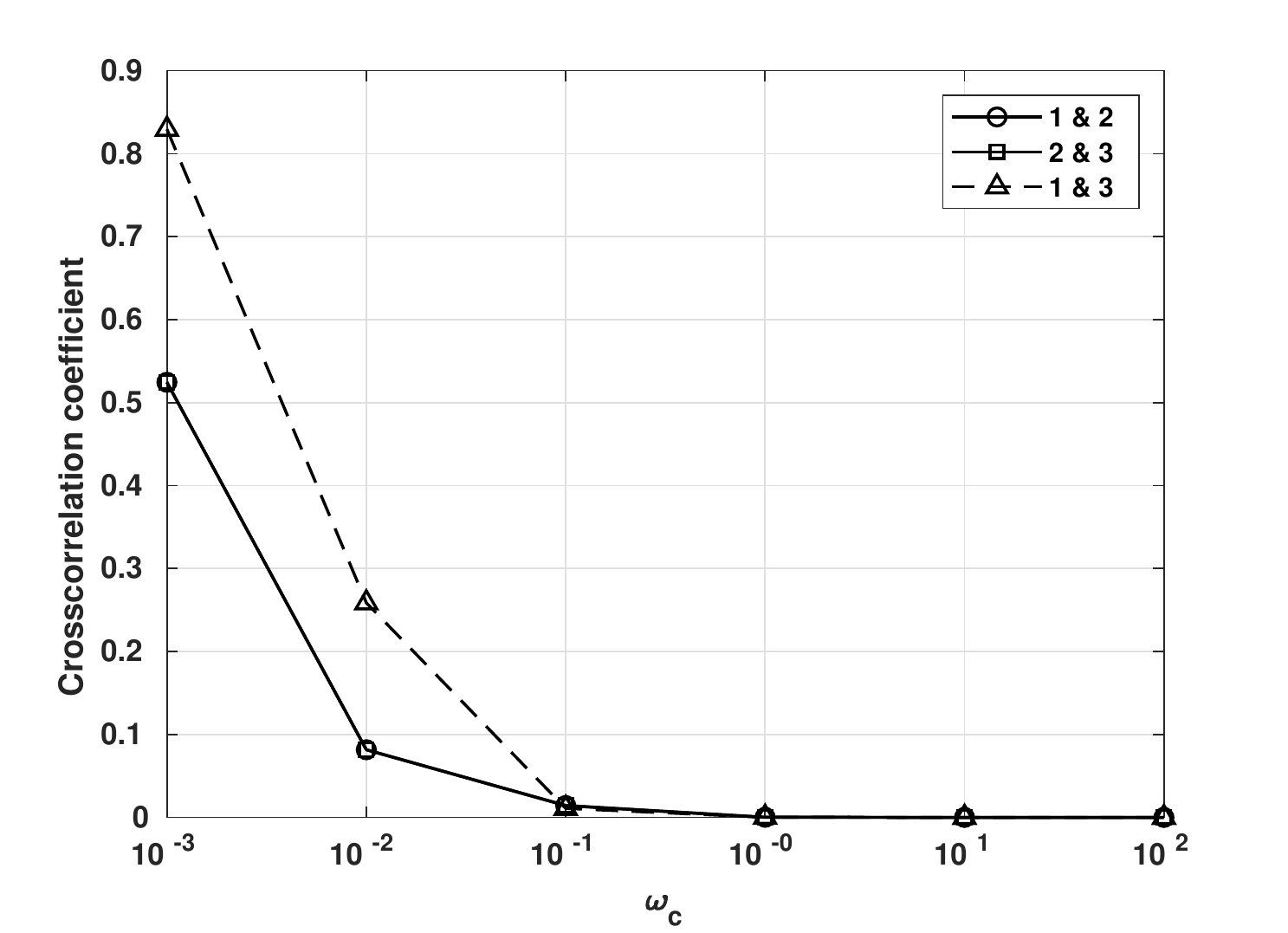}
		\caption{The comparison of the normalized magnitudes of the cross-correlation coefficients for three targets of interest at directions $\{-50\deg, 0\deg, 50\deg\}$ as functions of $\omega_c$.} 
		\label{fig:crosscorrelation_coefficients_wc}
	\end{figure}
	Herein, we compare the resulting beam-pattern with the desired one for the two cases of $\omega_c=1$ (with cross-correlation) and $\omega_c=0$ (without cross-correlation).
	Note that the designed beam-patterns obtained with and without considering the cross-correlation term are similar to one another.
	However, the cross-correlation behavior of the former is much better than that of the latter in that the probing signals corresponding to $ \omega_c = 1 $, are almost uncorrelated with each other.
	This can be further verified from Fig. \ref{fig:crosscorrelation_coefficients_wc}, where we provided the comparison of the normalized magnitudes of the cross-correlation coefficients (as formulated in the second term of the right hand side of \eqref{eq:J}) for the same three targets of interest at directions $\tilde{\theta} = \{-50\deg, 0\deg, 50\deg\}$, as functions of $\omega_c$.
	It is evident from Fig. \ref{fig:crosscorrelation_coefficients_wc} that when $\omega_c$ is very small (close to zero), the first and third reflected signals are highly correlated.
	On the other hand, for $\omega_c>0.1$ all cross-correlation coefficients are approximately zero. 
	The proposed algorithm outperforms the method in \cite{8378710} in terms of accuracy (measured in MSE), and additionally, is capable of designing waveform covariance matrices with low cross-correlation.
	\begin{figure}
		\centering
		\includegraphics[draft=false, width=0.4\textheight]{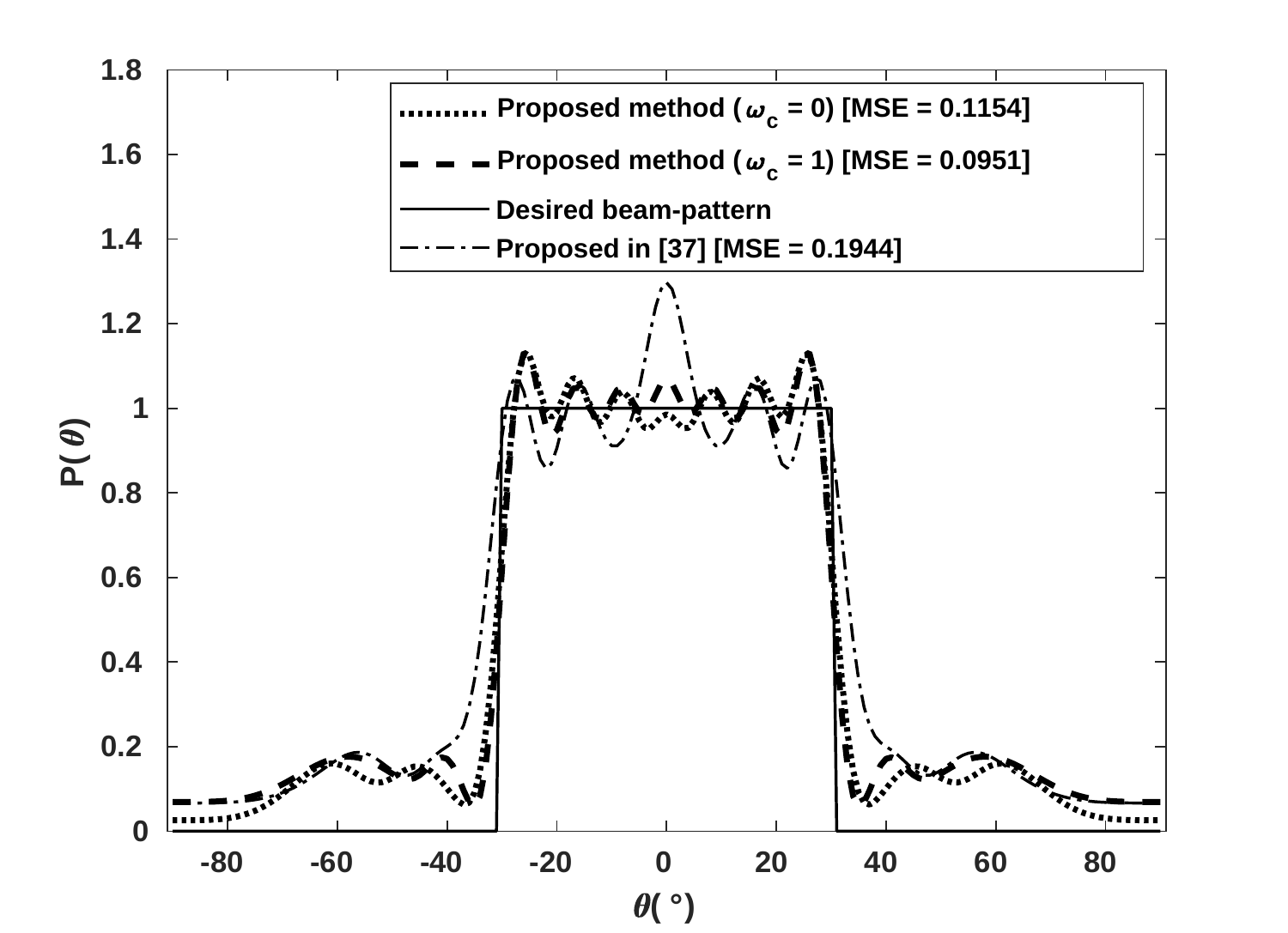}
		\caption{The transmit beam-pattern design for $ M = 15, N = 10 $ with and without the cross-correlation suppression with one mainlobe at $ \tilde{\theta}=0\deg $ with a beam-width of $ \triangle=60\deg $. It can be noted that our algorithm outperforms \cite{8378710}.} 
		\label{fig:beampattern_one_lobe}
	\end{figure}
	\begin{figure}
		\centering
		\includegraphics[draft=false, width=0.4\textheight]{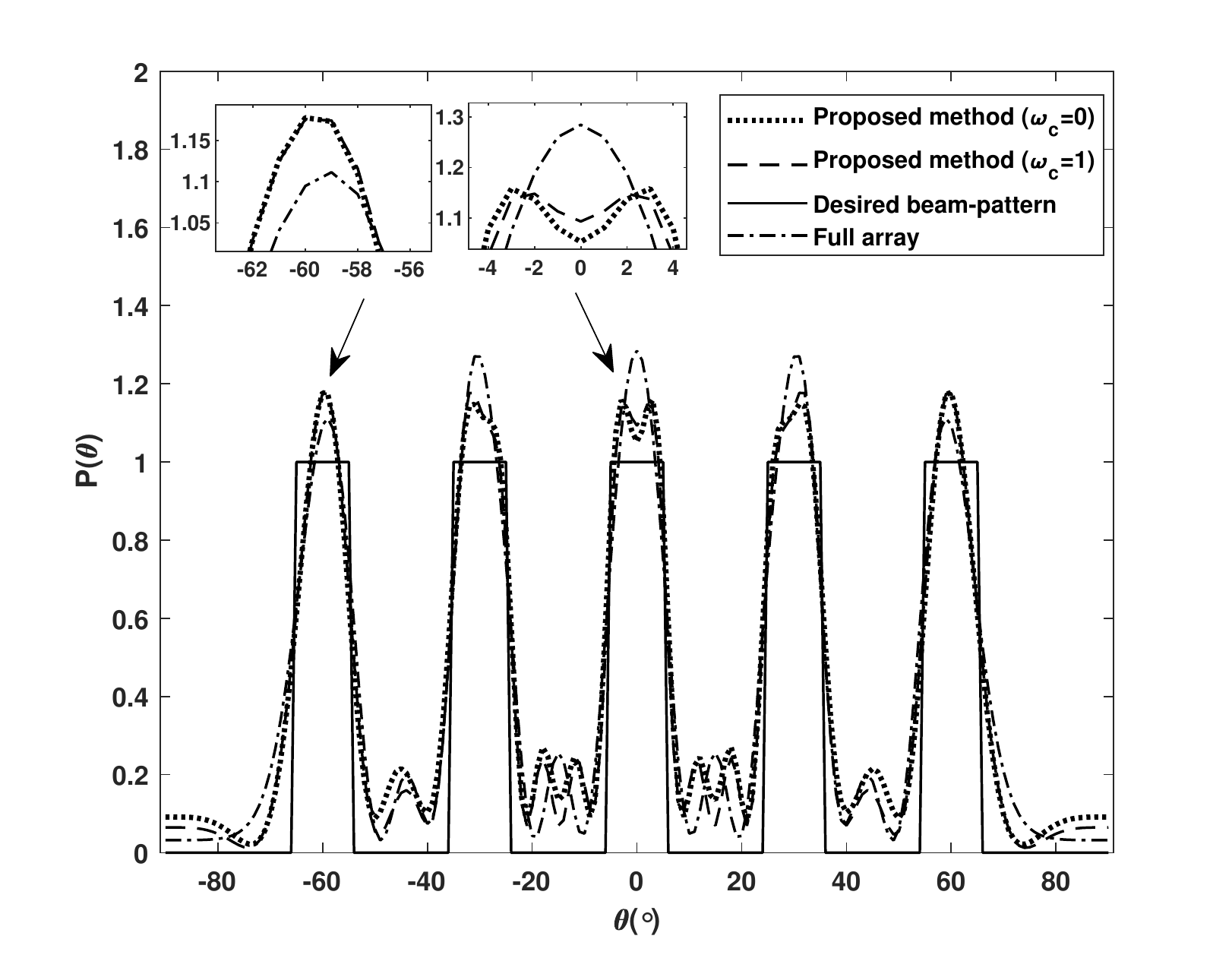}
		\caption{The transmit beam-pattern design for $ M = 20, N = 15 $ with and without the cross-correlation suppression with five mainlobes at $ \tilde{\theta} =	\{-60\deg, -30\deg, 0\deg, 30\deg, 60\deg\} $ with a beam-width of $ \triangle=10\deg$. Full array represents the $15$ antenna elements tightly placed in all $15$ grid points.} 
		\label{fig:beampattern_five_lobes}
	\end{figure}
	
	In Fig. \ref{fig:beampattern_one_lobe}, we further consider the design scenario of approximating the beam-patterns with one mainlobe at $\tilde{\theta} =  0\deg$, with a width of $ 60\deg$, with and without cross-correlation suppression.
	Note that in both cases of $\omega_c=0$ and $\omega_c=1$, our proposed method can accurately approximate the desired beam-pattern and provide a better beam-pattern than that of \cite{8378710}.
	
	Fig. \ref{fig:beampattern_five_lobes} shows the beam-pattern with five mainlobes at $\tilde{\theta} = \{-60\deg, -30\deg, 0\deg, 30\deg, 60\deg\}$ with a shorter beam-width of $10\deg$ for $M=20$ and $N=15$.
	We compare the beam-pattern approximated by our framework (\ie, with the configuration of 15 antennas placed in an array of 20 grid points) with that generated by a full linear array \ie $15$ antenna elements tightly placed in all $15$ grid points.
	It can be clearly seen from the Fig. \ref{fig:beampattern_five_lobes} that the proposed method approximates the beam-pattern better than that of full array.
	One can further notice that the transmitted power values are almost the same in all mainlobe despite being farther away from the central mainlobe, as compared to the full array.
	
	\begin{figure}
		\centering
		\includegraphics[draft=false, width=0.4\textheight]{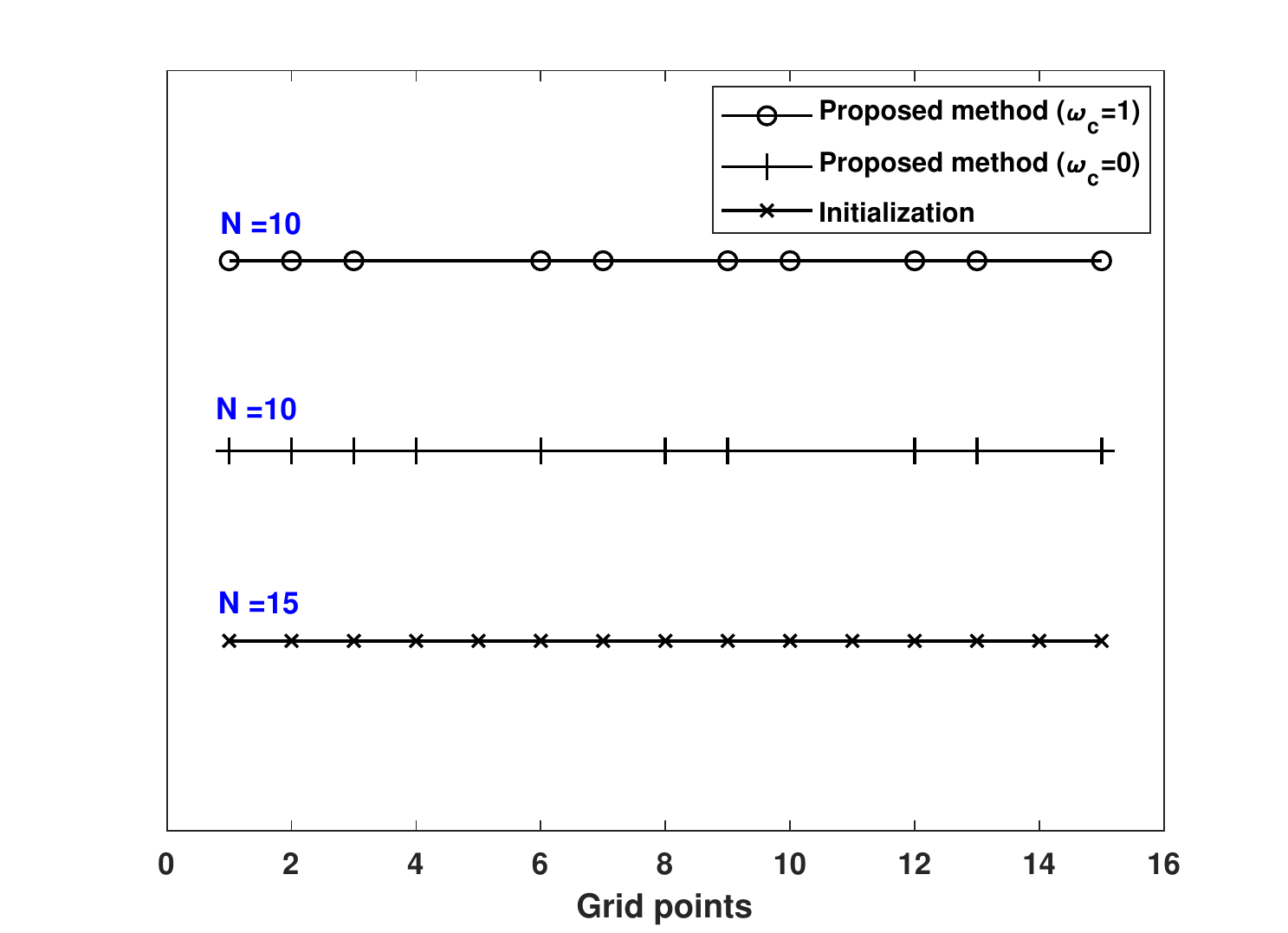}
		\caption{The antenna positions for $ M = 15, N = 10 $ with and without the correlation suppression.} 
		\label{fig:final_selected_antenna}
	\end{figure}

	In Fig. \ref{fig:final_selected_antenna}, we demonstrate the final antenna position vectors suggested by the proposed algorithm for the scenario considered in Fig. \ref{fig:beampattern_five_lobes} for the two cases of $ \omega_c = 0 $ and $ \omega_c = 1 $.
	It is interesting to note that the effective antenna aperture of the array is $M=15$, which can be safely reached by selecting only $N=10$ antennas.
	The corresponding beam-patterns are depicted in Fig. \ref{fig:beampattern_three_lobes}.
	
\begin{figure}
		\centering
		\includegraphics[draft=false, width=0.4\textheight]{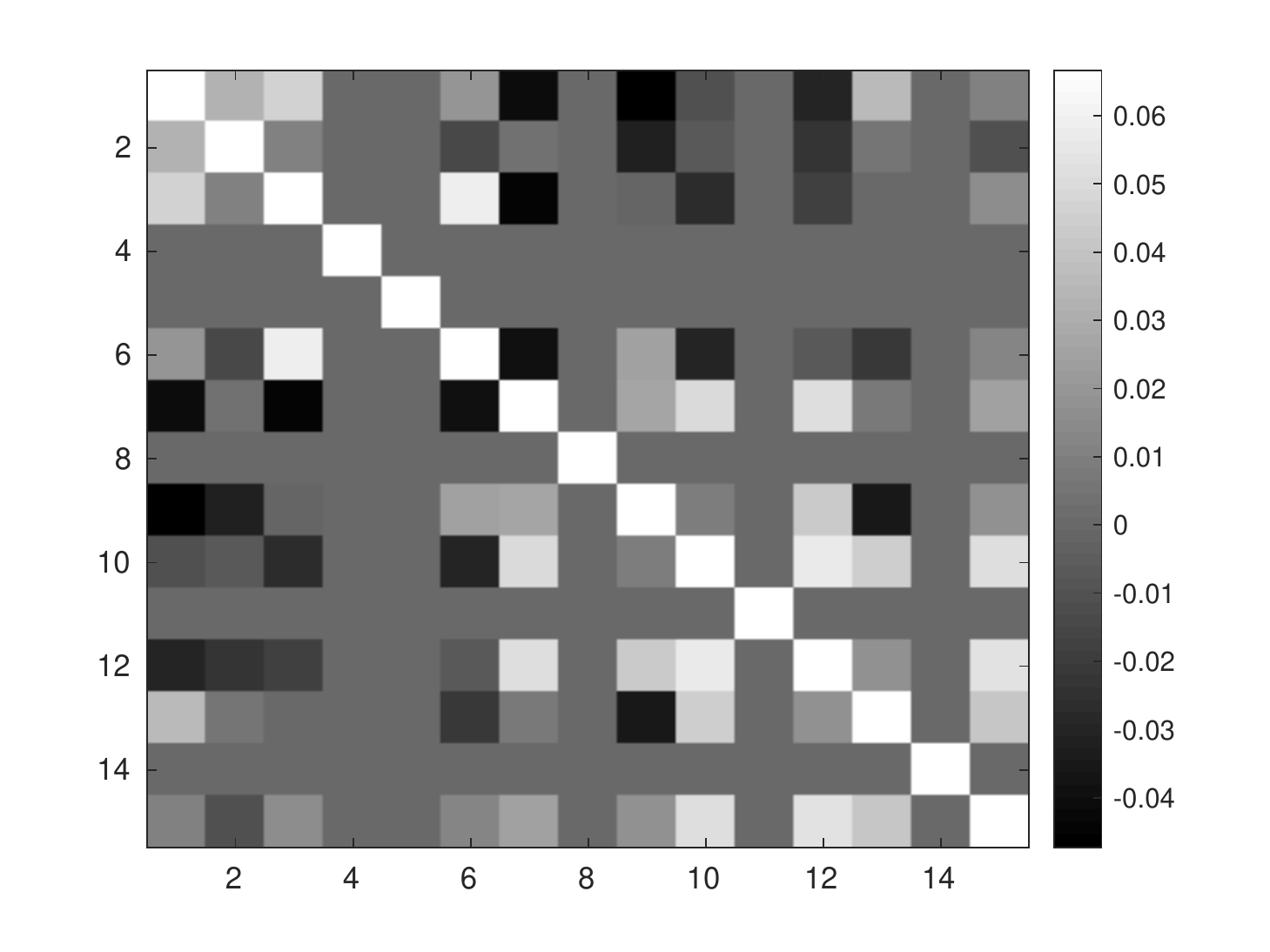}
		\caption{The generated $M\times M$ covariance matrix for $ M = 15, N = 10 $ with cross-correlation suppression.} 
		\label{fig:covariance_matrix_wc1}
	\end{figure}

	In addition, Fig. \ref{fig:covariance_matrix_wc1} shows the final $M\times M$ covariance matrix of the transmit signals, which can be used to design the transmitted sequence following specific requirements. 
	It can be readily shown that the generated matrix is symmetric and its eigenvalues are all non-negative (\ie, $\bR$ is a positive semidefinite matrix).
	It is interesting to note that the structure of the final covariance matrix is in agreement with the final antenna position vector, as reflected in the corresponding rows and columns of the rejected grid points, which are all zeros.

\begin{figure}
		\centering
		\includegraphics[draft=false, width=0.4\textheight]{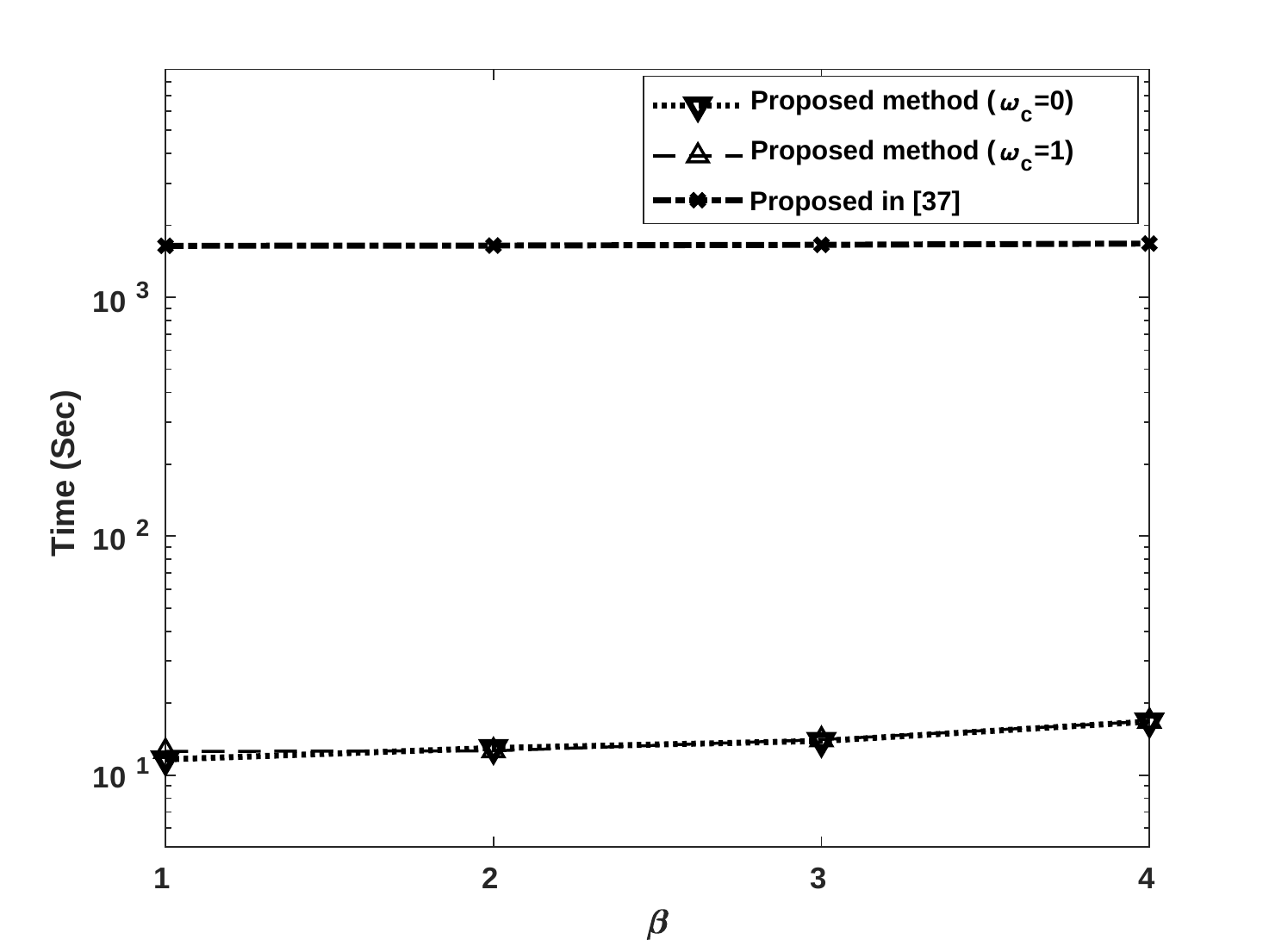}
		\caption{Comparison of the computational cost of our proposed algorithm and that of proposed in \cite{8378710} for different number of grid points and that of antennas. We consider $ M = 4 $ and $ N = 3 $ as initialization, and then linearly scale $ M $ and $ N $ by the factor of $ \beta= \{1, 2, 3, 4\} $.} 
		\label{fig:runtime_final}
	\end{figure}

	 In Fig. \ref{fig:runtime_final}, the comparison of the computational cost of the proposed algorithm and that of the method in \cite{8378710} for different number of grid points and antennas are shown.
	 For this experiment, we consider $M=4$ and $N=3$ as initialization, and then linearly scale $M$ and $N$ by the factor of $\beta \in\{1,2,3,4\}$.
	 The proposed algorithm significantly reduces the computational cost of the ADMM-based method in \cite{8378710} by \textit{a factor of more than $200$}.
	 To give it a perspective, \cite{8378710} takes $ \sim 4500 $ seconds to design the beam-pattern for $M=15$ and $N=10$, in around 20 outer iterations (on average) in a standard PC with 8-core processor and 16 GB memory.
	 Whereas, our proposed method finishes the same task in just 17 seconds using 3 outer iterations in the same standard PC, making the proposed framework particularly suitable for real-time applications.
	 
\begin{figure*}[t]
		\begin{minipage}[b]{0.48\linewidth}
			\centering
			\centerline{\includegraphics[draft=false,width=0.4\textheight]{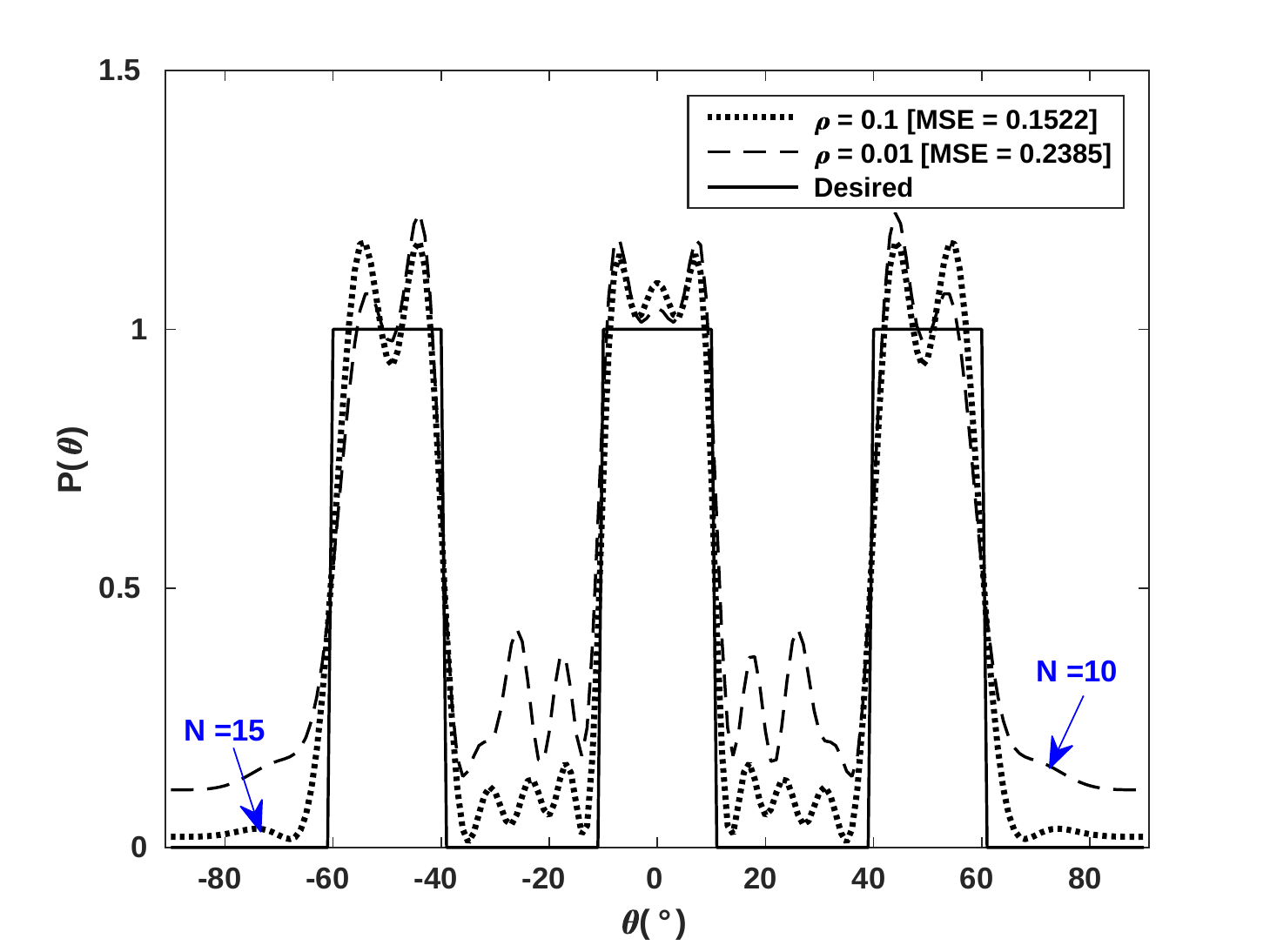}}  
			\centerline{(a)}\medskip
		\end{minipage}
		\hfill
		\begin{minipage}[b]{0.48\linewidth}
			\centering
			\centerline{\includegraphics[draft=false,width=0.4\textheight]{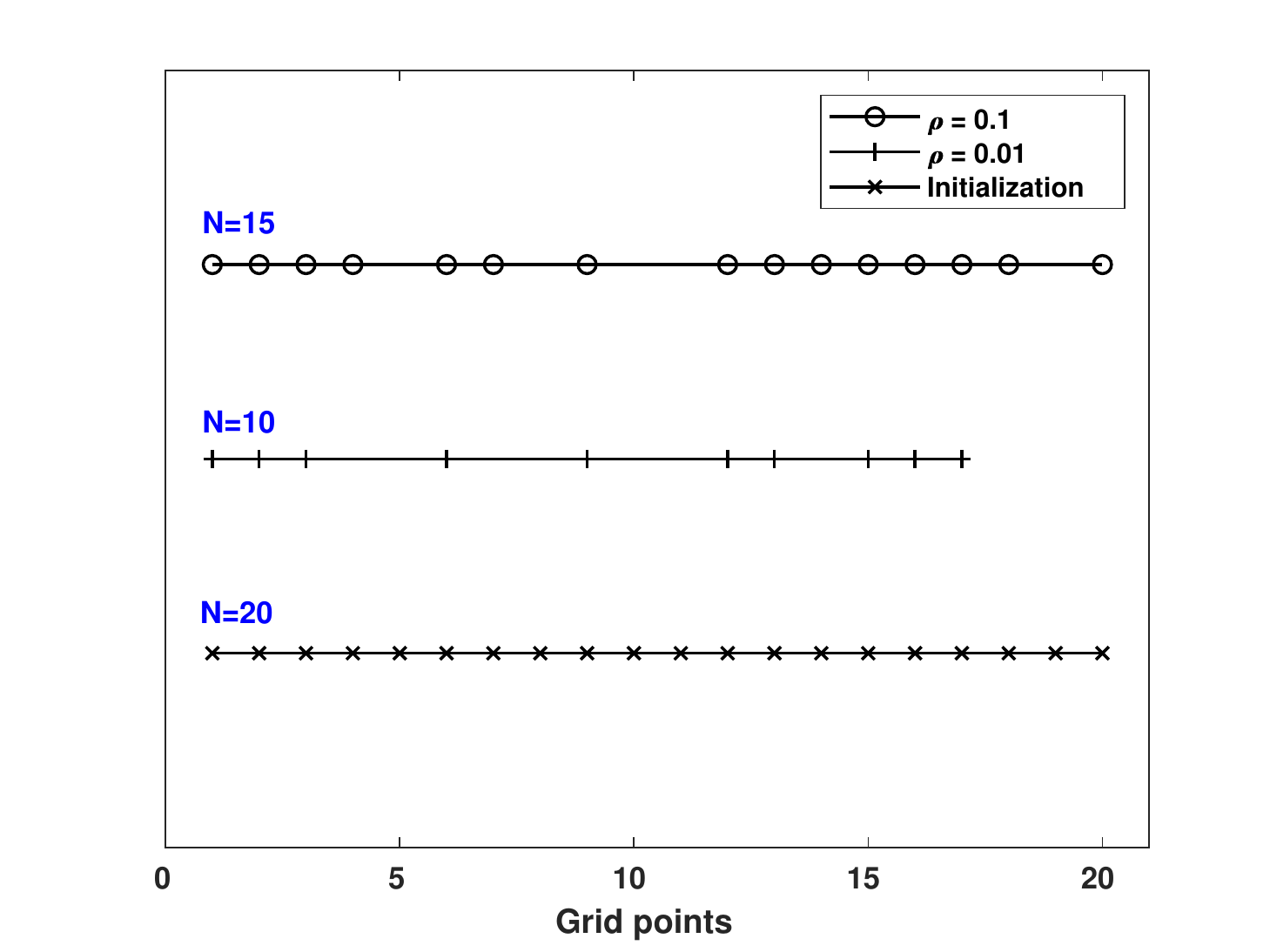}} 
			\centerline{(b)}\medskip
		\end{minipage}
		\caption{The (a) beam-pattern and, (b) final arrangements and total number of selected antennas for $M = 20$ grid points and penalty parameter $\rho = \{0.1, 0.01\}$.}
		\label{fig:generalized}
	\end{figure*}

	 Furthermore, Fig. \ref{fig:generalized} illustrates the beam-pattern design for the generalized case described in Section \ref{sec:exten}.
	 In the generalized case, we relax the constraints of \eqref{eq:probd} (\ie, $\|\bp\|_1=N$), and allow the total number of active antennas to deviate from $N$ (which can be chosen depending on the applications) via changing the penalty variable $\rho$.
	 For this simulation, we set $M=20$ and $N=15$ and provide the obtained beam-patterns and the final arrangement and total number of antennas suggested by the proposed algorithm, in Fig. \ref{fig:generalized}-(a) and \ref{fig:generalized}-(b), respectively for different values of $\rho$.
	 It is interesting to note that for $\rho=0.1$, the proposed algorithm successfully returns an arrangement with $15$ antennas as requested in the design parameter.
	 However, for $\rho=0.01$, the algorithm suggests an arrangement with $10$ antennas, which remains unchanged for $\rho<0.01$, suggesting $N=10$ is the minimum number of antenna that can be utilized. 
     Further note that the resulting beam-pattern for the two cases are similar in the mainlobes, although having different number of antennas. 
     
\begin{figure}
		\centering
		\includegraphics[draft=false, width=0.38\textheight]{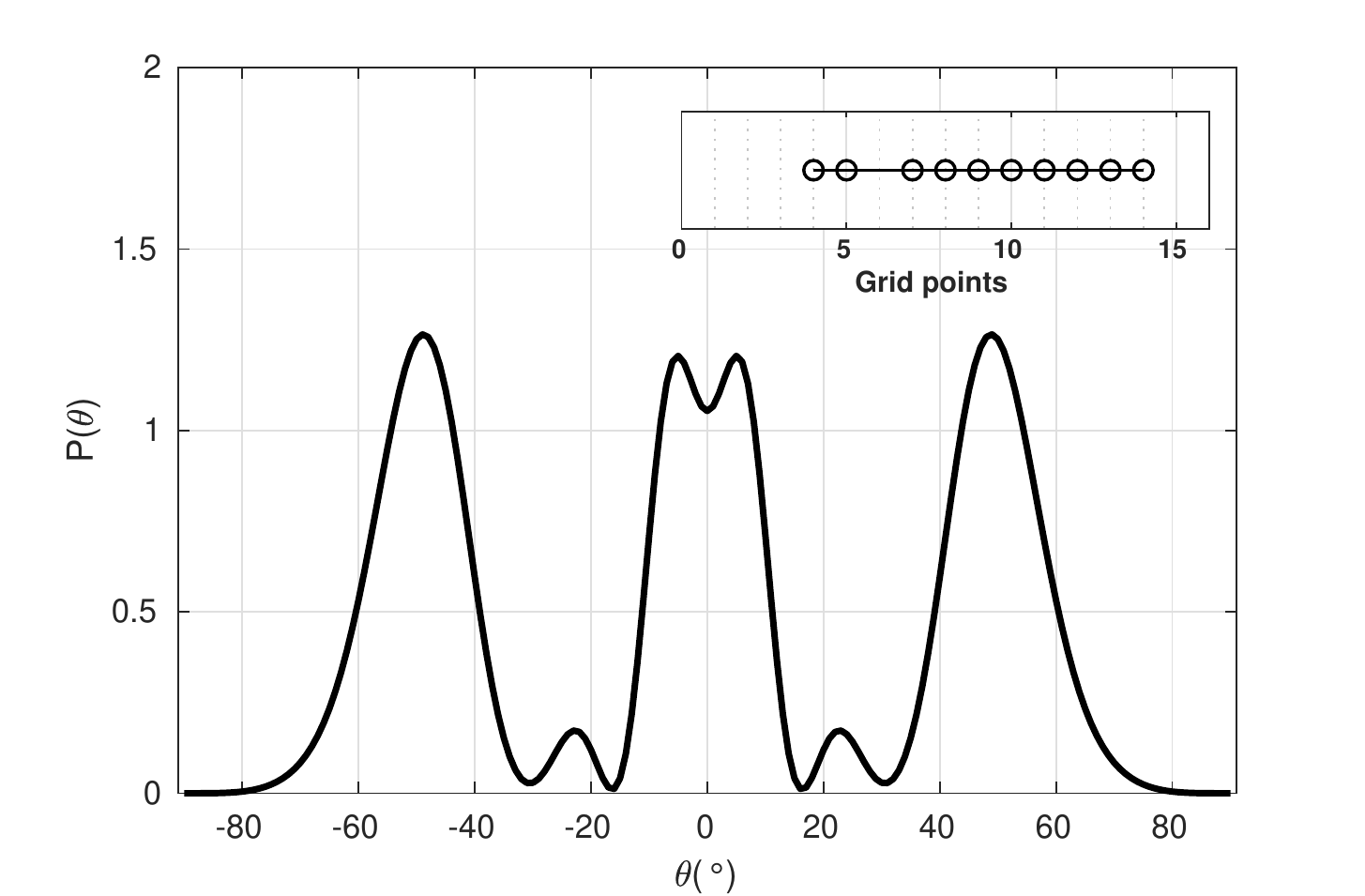}
		\caption{Convergence of the proposed algorithm: the approximated transmit beampattern and the optimized antenna positions (inset).} 
		\label{fig:conv_beampattern_three_lobes}
	\end{figure}

     In the next experiment, we further examine the convergence performance of the proposed algorithm.
     Especially, we perform a Monte-Carlo simulation ($n=1000$) with fixed parameters used in the experiment shown in Fig. \ref{fig:beampattern_three_lobes}.
     In each run, we initialize the waveform covariance matrix $\bR$ with a randomly generated PSD matrix while keeping all the other parameters unchanged.
     It is interesting to observe that, in each experiment, the designed beam pattern converges to the one shown in Fig. \ref{fig:conv_beampattern_three_lobes} for all $n=1000$.
     Furthermore, in each case, the optimized antenna positions are also the same as shown in Fig. \ref{fig:conv_beampattern_three_lobes} which implies that the proposed algorithm has satisfactory convergence performance.
     \nocite{khobahi2019model, milani2020intelligent}

\section{Conclusion} \label{sec:con}
	In this paper, the problem of jointly designing the probing signal covariance matrix as well as the antenna positions to approximate a given beam-pattern was studied.
	In order to tackle the problem, a novel cyclic (alternating) optimization method based on the non-convex formulation of the problem, was proposed.
	In addition, we used a greedy local search algorithm to tackle the non-convex problem of designing antenna position.
	Several numerical examples were provided which demonstrates the superiority of the proposed method over the existing methods in terms of accuracy and computational efficiency.

\appendices

\section{Time-complexity analysis of the proposed method}\label{app:b}
The computational complexity of the proposed method in Table \ref{tb:fullalgo} for a problem size of $(M,N)$ (\ie, $N$ antennas are to be selected from $M$ locations) can be obtained through the following steps:
\begin{enumerate}
	\item Evaluation of $\left(\bR^{(t)},\alpha^{(t)}\right)$ is a convex SDP problem and has a polynomial worst-case complexity \cite{vandenberghe1996semidefinite}.
	\item Evaluation of $\bp^{(t)}$ involves:
	\begin{enumerate}
		\item generation of $\bp_{\text{CS}}$, and calculation of $J(\bp)$ for each member of $\bp_{\text{CS}}$.
		\item choosing the best $\bp$.
	\end{enumerate}
\end{enumerate}
A careful investigation of the optimization step for $\bp$ reveals that it requires only $(M-N)$ inner-iterations (see Step 2 above).
Note that the generation of $\bp_{\text{CS}}$ and choosing the best $\bp$ (Step 2 above) linearly depend on the cardinality of the set $\bp_{\text{CS}}$, and thus, can be achieved in linear time-complexity.
Namely, assuming $|\bp_{\text{CS}}| = l$, the problem of finding $\bp^{*}\in\arg\min\;\bp_{CS}$ has a complexity of $\Oh{l}$.
For the $k$-th inner-iteration, let us denote the complexity of the calculations corresponding to the Step 2(a) above as a function of the cardinality of $\bp_{\text{CS}}$, \ie $\calC(l)$, where $l = M-k$.
Furthermore, the calculation of $J(\bp)$ for each $\bp$ has a constant cost $c$.
Hence, the total cost of optimization with respect to the vector $\bp$ at each outer-iteration admits the following upper-bound:
\begin{align}
\calC_{\text{tot}} &= \sum_{k=1}^{M-N}{c\cdot(M-k) + \calC(M-k)} \nonumber\\
&\leq \sum_{k=1}^{M-N}{cM + \calC(M)} \nonumber\\
&=(M-N)(cM) + (M-N)\calC(M) \nonumber \\
&\leq cM^2 + M\calC(M).
\end{align}
Thus, the worst-case complexity is $\Oh{M^2}$ (note that $\calC(M)$ corresponds to a complexity of $\Oh{M}$).

\bibliographystyle{IEEEbib}
\bibliography{refs}

\begin{thebibliography}{10}

\bibitem{asilomar2019bose}
A.~{Bose}, S.~{Khobahi}, and M.~{Soltanalian},
\newblock ``{Joint Optimization of Waveform Covariance Matrix and Antenna
  Selection for MIMO Radar},''
\newblock in {\em 2019 53rd Asilomar Conference on Signals, Systems, and
  Computers}, Nov 2019.

\bibitem{1316398}
E.~{Fishler}, A.~{Haimovich}, R.~{Blum}, D.~{Chizhik}, L.~{Cimini}, and
  R.~{Valenzuela},
\newblock ``{MIMO} radar: an idea whose time has come,''
\newblock in {\em Proceedings of the 2004 IEEE Radar Conference}, April 2004,
  pp. 71--78.

\bibitem{1399141}
F.~C. {Robey}, S.~{Coutts}, D.~{Weikle}, J.~C. {McHarg}, and K.~{Cuomo},
\newblock ``{MIMO} radar theory and experimental results,''
\newblock in {\em Conference Record of the Thirty-Eighth Asilomar Conference on
  Signals, Systems and Computers}, Nov 2004, vol.~1, pp. 300--304 Vol.1.

\bibitem{4516997}
D.~R. {Fuhrmann} and G.~{San Antonio},
\newblock ``Transmit beamforming for {MIMO} radar systems using signal
  cross-correlation,''
\newblock {\em IEEE Transactions on Aerospace and Electronic Systems}, vol. 44,
  no. 1, pp. 171--186, January 2008.

\bibitem{4350230}
J.~{Li} and P.~{Stoica},
\newblock ``{MIMO} radar with colocated antennas,''
\newblock {\em IEEE Signal Processing Magazine}, vol. 24, no. 5, pp. 106--114,
  Sep. 2007.

\bibitem{4358016}
J.~{Li}, P.~{Stoica}, L.~{Xu}, and W.~{Roberts},
\newblock ``On parameter identifiability of {MIMO} radar,''
\newblock {\em IEEE Signal Processing Letters}, vol. 14, no. 12, pp. 968--971,
  Dec 2007.

\bibitem{li2008mimo}
J.~Li and P.~Stoica,
\newblock {\em {MIMO} Radar Signal Processing},
\newblock Wiley - IEEE. Wiley, 2008.

\bibitem{4408448}
A.~M. {Haimovich}, R.~S. {Blum}, and L.~J. {Cimini},
\newblock ``{MIMO} radar with widely separated antennas,''
\newblock {\em IEEE Signal Processing Magazine}, vol. 25, no. 1, pp. 116--129,
  2008.

\bibitem{khobahi2019deepdararwave}
S.~Khobahi, A.~Bose, and M.~Soltanalian,
\newblock ``Deep radar waveform design for efficient automotive radar
  sensing,''
\newblock {\em arXiv preprint arXiv:1912.08180}, 2019.

\bibitem{1599974}
K.~W. {Forsythe} and D.~W. {Bliss},
\newblock ``Waveform correlation and optimization issues for {MIMO} radar,''
\newblock in {\em Conference Record of the Thirty-Ninth Asilomar Conference
  onSignals, Systems and Computers, 2005.}, Oct 2005, pp. 1306--1310.

\bibitem{1291865}
D.~W. {Bliss} and K.~W. {Forsythe},
\newblock ``Multiple-input multiple-output ({MIMO}) radar and imaging: degrees
  of freedom and resolution,''
\newblock in {\em The Thrity-Seventh Asilomar Conference on Signals, Systems
  Computers, 2003}, Nov 2003, vol.~1, pp. 54--59 Vol.1.

\bibitem{4176505}
N.~H. {Lehmann}, A.~M. {Haimovich}, R.~S. {Blum}, and L.~{Cimini},
\newblock ``High resolution capabilities of {MIMO} radar,''
\newblock in {\em 2006 Fortieth Asilomar Conference on Signals, Systems and
  Computers}, Oct 2006, pp. 25--30.

\bibitem{1597550}
E.~{Fishler}, A.~{Haimovich}, R.~S. {Blum}, L.~J. {Cimini}, D.~{Chizhik}, and
  R.~A. {Valenzuela},
\newblock ``Spatial diversity in radars—models and detection performance,''
\newblock {\em IEEE Transactions on Signal Processing}, vol. 54, no. 3, pp.
  823--838, March 2006.

\bibitem{5466526}
H.~{Godrich}, A.~M. {Haimovich}, and R.~S. {Blum},
\newblock ``Target localization accuracy gain in {MIMO} radar-based systems,''
\newblock {\em IEEE Transactions on Information Theory}, vol. 56, no. 6, pp.
  2783--2803, June 2010.

\bibitem{5393291}
Q.~{He}, R.~S. {Blum}, H.~{Godrich}, and A.~M. {Haimovich},
\newblock ``Target velocity estimation and antenna placement for {MIMO} radar
  with widely separated antennas,''
\newblock {\em IEEE Journal of Selected Topics in Signal Processing}, vol. 4,
  no. 1, pp. 79--100, Feb 2010.

\bibitem{1703855}
I.~{Bekkerman} and J.~{Tabrikian},
\newblock ``Target detection and localization using {MIMO} radars and sonars,''
\newblock {\em IEEE Transactions on Signal Processing}, vol. 54, no. 10, pp.
  3873--3883, Oct 2006.

\bibitem{khobahi2018optimized}
S.~{Khobahi} and M.~{Soltanalian},
\newblock ``Optimized transmission for consensus in wireless sensor networks,''
\newblock in {\em 2018 IEEE International Conference on Acoustics, Speech and
  Signal Processing (ICASSP)}, April 2018, pp. 3419--3423.

\bibitem{6649991}
G.~{Cui}, H.~{Li}, and M.~{Rangaswamy},
\newblock ``{MIMO} radar waveform design with constant modulus and similarity
  constraints,''
\newblock {\em IEEE Transactions on Signal Processing}, vol. 62, no. 2, pp.
  343--353, Jan 2014.

\bibitem{8141978}
Z.~{Cheng}, Z.~{He}, B.~{Liao}, and M.~{Fang},
\newblock ``{MIMO} radar waveform design with papr and similarity
  constraints,''
\newblock {\em IEEE Transactions on Signal Processing}, vol. 66, no. 4, pp.
  968--981, Feb 2018.

\bibitem{6472022}
M.~{Soltanalian}, B.~{Tang}, J.~{Li}, and P.~{Stoica},
\newblock ``Joint design of the receive filter and transmit sequence for active
  sensing,''
\newblock {\em IEEE Signal Processing Letters}, vol. 20, no. 5, pp. 423--426,
  May 2013.

\bibitem{4524058}
P.~{Stoica}, J.~{Li}, and X.~{Zhu},
\newblock ``Waveform synthesis for diversity-based transmit beampattern
  design,''
\newblock {\em IEEE Transactions on Signal Processing}, vol. 56, no. 6, pp.
  2593--2598, June 2008.

\bibitem{4567663}
J.~{Li}, P.~{Stoica}, and X.~{Zheng},
\newblock ``Signal synthesis and receiver design for {MIMO} radar imaging,''
\newblock {\em IEEE Transactions on Signal Processing}, vol. 56, no. 8, pp.
  3959--3968, Aug 2008.

\bibitem{7811203}
H.~{Li}, Y.~{Zhao}, Z.~{Cheng}, and D.~{Feng},
\newblock ``Correlated {LFM} waveform set design for {MIMO} radar transmit
  beampattern,''
\newblock {\em IEEE Geoscience and Remote Sensing Letters}, vol. 14, no. 3, pp.
  329--333, March 2017.

\bibitem{soltanalian2014single}
M.~Soltanalian, H.~Hu, and P.~Stoica,
\newblock ``Single-stage transmit beamforming design for {MIMO} radar,''
\newblock {\em Signal Processing}, vol. 102, pp. 132 -- 138, 2014.

\bibitem{4276989}
P.~Stoica, J.~Li, and Y.~Xie,
\newblock ``On probing signal design for {MIMO} radar,''
\newblock {\em IEEE Transactions on Signal Processing}, vol. 55, no. 8, pp.
  4151--4161, Aug 2007.

\bibitem{aa1}
A.~{Ameri}, J.~{Li}, and M.~{Soltanalian},
\newblock ``One-bit radar processing and estimation with time-varying sampling
  thresholds,''
\newblock in {\em 2018 IEEE 10th Sensor Array and Multichannel Signal
  Processing Workshop (SAM)}, July 2018, pp. 208--212.

\bibitem{6747391}
J.~{Lipor}, S.~{Ahmed}, and M.~{Alouini},
\newblock ``Fourier-based transmit beampattern design using {MIMO} radar,''
\newblock {\em IEEE Transactions on Signal Processing}, vol. 62, no. 9, pp.
  2226--2235, May 2014.

\bibitem{7178393}
T.~{Bouchoucha}, S.~{Ahmed}, T.~Y. {Al-Naffouri}, and M.~{Alouini},
\newblock ``Closed-form solution to directly design face waveforms for
  beampatterns using planar array,''
\newblock in {\em 2015 IEEE International Conference on Acoustics, Speech and
  Signal Processing (ICASSP)}, April 2015, pp. 2359--2363.

\bibitem{7829401}
T.~{Bouchoucha}, S.~{Ahmed}, T.~Y. {Al-Naffouri}, and M.~{Alouini},
\newblock ``{DFT}-based closed-form covariance matrix and direct waveforms
  design for {MIMO} radar to achieve desired beampatterns,''
\newblock {\em IEEE Transactions on Signal Processing}, vol. 65, no. 8, pp.
  2104--2113, April 2017.

\bibitem{khobahi2018signal}
S.~{Khobahi} and M.~{Soltanalian},
\newblock ``Signal recovery from 1-bit quantized noisy samples via adaptive
  thresholding,''
\newblock in {\em 2018 52nd Asilomar Conference on Signals, Systems, and
  Computers}, Oct 2018, pp. 1757--1761.

\bibitem{5765721}
S.~{Ahmed}, J.~S. {Thompson}, Y.~R. {Petillot}, and B.~{Mulgrew},
\newblock ``Unconstrained synthesis of covariance matrix for {MIMO} radar
  transmit beampattern,''
\newblock {\em IEEE Transactions on Signal Processing}, vol. 59, no. 8, pp.
  3837--3849, Aug 2011.

\bibitem{5962371}
S.~{Ahmed}, J.~S. {Thompson}, Y.~R. {Petillot}, and B.~{Mulgrew},
\newblock ``Finite alphabet constant-envelope waveform design for {MIMO}
  radar,''
\newblock {\em IEEE Transactions on Signal Processing}, vol. 59, no. 11, pp.
  5326--5337, Nov 2011.

\bibitem{7915123}
Z.~{Cheng}, Z.~{He}, R.~{Li}, and Z.~{Wang},
\newblock ``Robust transmit beampattern matching synthesis for {MIMO} radar,''
\newblock {\em Electronics Letters}, vol. 53, no. 9, pp. 620--622, 2017.

\bibitem{6698378}
M.~{Soltanalian} and P.~{Stoica},
\newblock ``Designing unimodular codes via quadratic optimization,''
\newblock {\em IEEE Transactions on Signal Processing}, vol. 62, no. 5, pp.
  1221--1234, March 2014.

\bibitem{7955071}
Z.~{Cheng}, Z.~{He}, S.~{Zhang}, and J.~{Li},
\newblock ``Constant modulus waveform design for {MIMO} radar transmit
  beampattern,''
\newblock {\em IEEE Transactions on Signal Processing}, vol. 65, no. 18, pp.
  4912--4923, Sep. 2017.

\bibitem{7027831}
X.~{Zhang}, Z.~{He}, L.~{Rayman-Bacchus}, and J.~{Yan},
\newblock ``{MIMO} radar transmit beampattern matching design,''
\newblock {\em IEEE Transactions on Signal Processing}, vol. 63, no. 8, pp.
  2049--2056, April 2015.

\bibitem{8378710}
Z.~{Cheng}, Y.~{Lu}, Z.~{He}, , J.~{Li}, and X.~{Luo},
\newblock ``Joint optimization of covariance matrix and antenna position for
  {MIMO} radar transmit beampattern matching design,''
\newblock in {\em 2018 IEEE Radar Conference (RadarConf18)}, April 2018, pp.
  1073--1077.

\bibitem{boyd2011distributed}
S.~Boyd, N.~Parikh, E.~Chu, B.~Peleato, J.~Eckstein, et~al.,
\newblock ``Distributed optimization and statistical learning via the
  alternating direction method of multipliers,''
\newblock {\em Foundations and Trends{\textregistered} in Machine learning},
  vol. 3, no. 1, pp. 1--122, 2011.

\bibitem{8683876}
S.~{Khobahi}, N.~{Naimipour}, M.~{Soltanalian}, and Y.~C. {Eldar},
\newblock ``Deep signal recovery with one-bit quantization,''
\newblock in {\em ICASSP 2019 - 2019 IEEE International Conference on
  Acoustics, Speech and Signal Processing (ICASSP)}, May 2019, pp. 2987--2991.

\bibitem{nocedal2006numerical}
J.~Nocedal and S.~Wright,
\newblock {\em Numerical optimization},
\newblock Springer Science \& Business Media, 2006.

\bibitem{DaRonco2014}
Claudio~Comis Da~Ronco and Ernesto Benini,
\newblock {\em A Simplex-Crossover-Based Multi-Objective Evolutionary
  Algorithm}, pp. 583--598,
\newblock Springer Netherlands, Dordrecht, 2014.

\bibitem{grant2014cvx}
M.~Grant and S.~Boyd,
\newblock ``{CVX}: Matlab software for disciplined convex programming, version
  2.1,'' \url{http://cvxr.com/cvx}, Mar. 2014.

\bibitem{khobahi2019model}
S.~Khobahi and M.~Soltanalian,
\newblock ``Model-aware deep architectures for one-bit compressive variational
  autoencoding,''
\newblock {\em arXiv preprint arXiv:1911.12410}, 2019.

\bibitem{milani2020intelligent}
O.~H. Milani, S.~A. Motamedi, and S.~Sharifian,
\newblock ``Intelligent service selection in a multi-dimensional environment of
  cloud providers for iot stream data through cloudlets,'' 2020.

\bibitem{vandenberghe1996semidefinite}
L.~Vandenberghe and S.~Boyd,
\newblock ``Semidefinite programming,''
\newblock {\em SIAM Review}, vol. 38, no. 1, pp. 49--95, 1996.

\end{thebibliography}
\end{document}